\documentclass[12pt,preprint]{aastex}
\begin{document}
\def\fas{\hbox{$.\!\!''$}}
\def\Msun{$M_{\odot}$}
\def\Rsun{$R_{\odot}$}
\def\Msunyr{$M_{\odot}~yr^{-1}$}
\def\flux{ergs~cm$^{-2}$~s$^{-1}$}
\def\fluxlam{ergs s$^{-1}$ cm$^{-2}$\AA$^{-1}$}
\def\lum{ergs~s$^{-1}$}
\def\deg{$^{\rm o}$}

\shorttitle{Spectrum of Black Hole XTE J1118+480}
\shortauthors{McClintock et al.}

\title{Multiwavelength Spectrum of the Black Hole XTE~J1118+480 in
Quiescence\altaffilmark{1,2}}

\author{Jeffrey E. McClintock\altaffilmark{3}, Ramesh
Narayan\altaffilmark{3}, Michael R. Garcia\altaffilmark{3}, Jerome
A. Orosz\altaffilmark{4}, \newline Ronald
A. Remillard\altaffilmark{5}, Stephen S. Murray\altaffilmark{3}}

\altaffiltext{1}{Based on observations with the
NASA/ESA {\it Hubble Space Telescope} obtained at the Space Telescope
Science Institute, which is operated by the Association of
Universities for Research in Astronomy, Inc., under NASA contract
NAS5-26555.}

\altaffiltext{2}{Observations reported here were obtained at the MMT
Observatory, a facility operated jointly by the University of Arizona
and the Smithsonian Institution.}

\altaffiltext{3}{Harvard-Smithsonian Center for Astrophysics, 60
Garden Street, Cambridge, MA 02138; jmcclintock@cfa.harvard.edu,
rnarayan@cfa.harvard.edu, mgarcia@cfa.harvard.edu,
smurray@cfa.harvard.edu}

\altaffiltext{4}{Department of Astronomy, San Diego State University,
5500 Campanile Drive, San Diego, CA 92182; orosz@zwartgat.sdsu.edu}

\altaffiltext{5}{Center for Space Research, MIT, Cambridge, MA 02139;
rr@space.mit.edu}

\begin{abstract}

We present an X--ray/UV/optical spectrum of the black hole primary in
the X--ray nova XTE~J1118+480 in quiescence at $L_{\rm
x}~\approx~4~\times~10^{-9}L_{\rm Edd}$.  The {\it Chandra}, {\it HST}
and MMT spectroscopic observations were performed simultaneously on
2002 January~12~UT.  Because this 4.1-hr binary is located at
b~=~$62^{\rm o}$, the transmission of the ISM is very high (e.g., 70\%
at 0.3 keV).  We present many new results for the quiescent state,
such as the first far--UV spectrum and evidence for an 0.35 mag
orbital modulation in the near--UV flux.  However, the centerpiece of
our work is the multiwavelength spectrum of XTE~J1118+480, which we
argue represents the canonical spectrum of a stellar--mass black hole
radiating at $L_{\rm x}~\sim~10^{-8.5}L_{\rm Edd}$.  This spectrum is
comprised of two apparently disjoint components: a hard X--ray
spectrum with a photon index $\Gamma~=~2.02~\pm~0.16$, and an
optical/UV continuum that resembles a 13,000~K disk blackbody spectrum
punctuated by several strong emission lines.  We present a model of
the source in which the accretion flow has two components: (1) an
X--ray--emitting interior region where the flow is
advection--dominated, and (2) a thin, exterior accretion disk with a
truncated inner edge ($R_{\rm tr}~\sim~10^{4}$~Schwarzschild radii)
that is responsible for the optical/UV spectrum. For $D~=~1.8$~kpc,
the luminosity of the X--ray component is $L_{\rm
x}~\approx~3.5~\times~10^{30}$~\lum~(0.3--7~keV); the bolometric
luminosity of the optical/UV component is $\approx~20$~times greater.

\end{abstract}


\section{Introduction}

An X--ray nova (a.k.a. soft X--ray transient) typically brightens in
X--rays by as much as $10^{7}$ in a week and then decays back into
quiescence over the course of a year (van Paradijs \& McClintock
1995).  The X--ray nova XTE J1118+480 (hereafter J1118) was discovered
on 2000 March 29 (Remillard et al. 2000).  Its extraordinarily high
Galactic latitude (b~=~+62\deg) and correspondingly low interstellar
absorption ($N_{\rm H}~\approx~1.2~\times~10^{20}$~cm$^{-2}$) provide a
unique opportunity to probe the soft X--ray and ultraviolet spectrum of
a stellar--mass black hole. For this reason, the multiwavelength
spectrum of J1118 was closely observed during its outburst in 2000
(Hynes et al. 2000; McClintock et al. 2001b; Frontera et al. 2001).
Throughout its outburst, the source remained in the low/hard state,
one of the characteristic spectral states of an accreting black hole
binary.  Associated radio emission was detected and interpreted in
terms of a steady radio jet (Fender et al. 2001).  The
optical/UV/X--ray spectrum clearly exhibited two components, one of
which was interpreted as thermal emission from a truncated accretion
disk (Esin et al. 2001).

By late October of 2000, J1118 was nearly quiescent. Shortly
thereafter it was shown that the mass of the compact primary is
definitely greater than 6~\Msun, thereby establishing that the compact
primary is a black hole (McClintock et al. 2001a; Wagner et al. 2001).
The orbital period is the shortest known for a black hole X--ray nova,
4.1~hr, the spectral type of the secondary is approximately K5, and
the inclination of the system is high, $i$~$\sim~80$\deg.  Here we
report on the broadband energy distribution of this black hole binary
in its quiescent state (Zurita et al. 2002).  Our results are based on
observations that were made simultaneously on 2002 January 12 (UT)
using the {\it Chandra X--ray Observatory (CXO)}, the {\it Hubble
Space Telescope (HST)} and the 6.5m MMT telescope.

The quiescent X--ray luminosities of short-period X--ray novae like
J1118 are known to be extraordinarily low,
$L_{\rm x}~\sim~10^{31}$\lum~$\sim~10^{-8.5}L_{\rm Edd}$ (McClintock,
Horne \& Remillard 1995; Garcia et al. 2001), and therefore the
quality of the spectral data is severely limited by counting
statistics.  Nevertheless, we undertook these observations in order to
make crucial tests of the advection--dominated accretion flow (ADAF)
model (Narayan \& Yi 1994, 1995b; Abramowicz et al. 1995; see Narayan,
Mahadevan \& Quataert 1998 for a review).  The plain physics behind
this widely-used model is one of its chief virtues: It describes black
hole accretion in terms of a hot two-temperature plasma and familiar
radiation physics such as synchrotron/bremsstrahlung emission and
Compton scattering.

At low mass accretion rates there is good evidence that the accretion
flow is optically thin and dominated by advection.  In this regime,
ADAF models provide a satisfactory description of the observations.
Some successes of the ADAF model include the prediction, subsequently
confirmed by observations, that the accretion disk is truncated at a
large inner radius in black hole X-ray binaries in the low/hard and
quiescent spectral states (Narayan 1996; Esin, McClintock \& Narayan
1997), an explanation for the spectrum of black hole X--ray novae in
quiescence (Narayan, McClintock \& Yi 1996; Narayan, Barret \&
McClintock 1997a), the prediction and confirmation that at low mass
accretion rates black holes are very much fainter than neutron stars
(Narayan, Garcia \& McClintock 1997b, 2002; Garcia et al. 2001;
Hameury et al. 2002), and an explanation for the delay in the rise of
the X--ray light curve relative to the optical and UV light curves
when X--ray novae go into outburst (Hameury et al. 1997).  The ADAF
model also provides a natural explanation for why Sgr~A$^{*}$ in our
Galactic Center as well as the nuclei of most external galaxies are
far dimmer in X-rays than one would predict if one assumed accretion at the
Bondi-Hoyle rate feeding a standard thin accretion disk.  The new
paradigm is that these systems all accrete via radiatively inefficient
ADAF-like flows (Narayan, Yi, \& Mahadevan 1995; Fabian \& Rees 1995;
Lasota et al. 1996; Quataert et al. 1999; Di~Matteo et al. 2000, 2001,
2002; see Narayan 2002 for a review).

In the context of the ADAF model, there is no distinction between the
quiescent state and the low/hard state of a black hole binary, except
that the mass transfer rate/luminosity is much higher for the latter
state (Narayan 1996; Esin et al. 1997).  Moreover, observationally
there is no evidence for a transition between the two states that
might signal a reconfiguration of the accretion flow (such as the
transition between the low/hard and high/soft states; e.g., Esin et
al. 1997).  In both the low/hard and quiescent states, the ADAF model
predicts qualitatively that the inner edge of the accretion disk is
truncated at some large radius, with the interior region filled by an
ADAF.  Strong evidence for such a truncated disk with an inner radius
of $\gtrsim~55$~Schwarzschild radii and a hot, optically--thin plasma
in the interior region was provided by observations of J1118 in the
low/hard state during outburst (McClintock et al. 2001b; Esin et
al. 2001).  With the system now in quiescence, the ADAF model predicts
that the inner disk edge will have moved further outward (Esin et
al. 1997).

ADAF models have been fitted to the quiescent X--ray/optical continuum
spectra of A0620--00, V404 Cyg and GRO J1655--40 (Narayan et al. 1996;
Narayan et al. 1997a; Hameury et al. 1997; Quataert \& Narayan 1999).
However, because the quiescent state can be quite variable (Wagner et
al. 1994; Garcia et al. 2001), this work has suffered from the fact
that the X--ray and optical data were not obtained simultaneously.  A
multiwavelength spectrum of A0620--00 that includes the near--UV (NUV)
band (2000--3000~\AA) was reported by McClintock \& Remillard (2000).
Again, however, the data in the separate wavebands was not obtained
simultaneously; furthermore, the sizable and uncertain extinction
corrections were a limitation. In this regard, we stress that
absorption by the ISM is almost negligible for J1118: For our adopted
column depth of $N_{\rm H}~=~1.2~\times~10^{20}$~cm$^{-2}$, the
transmission of the ISM is 70\% for the softest X--rays (0.3 keV)
considered herein (Balucinska-Church \& McCammon 1992).  Similarly,
the minimum transmission in the UV for the wavelengths considered here
($\geq~1320$~\AA) is 82\% (Predehl \& Schmitt 1995; Cardelli, Clayton,
\& Mathis 1989).

In this paper we present a high--quality spectrum of an accreting
stellar--mass black hole in quiescence ($L_{\rm x}~\sim~10^{-8.5}L_{\rm
Edd}$).  {\it There are two virtues of this spectrum of J1118 that set
it apart from all previous multiwavelength spectra obtained for
quiescent binary black holes: The optical, near--UV and X--ray data
reported on herein were obtained simultaneously, and absorption by the
ISM is practically negligible.}

This work is organized as follows.  In \S2 we stress the simultaneity
of the observations and then discuss in turn the data collection and
analysis of the {\it Chandra} X--ray data, the {\it HST} ultraviolet
data and the MMT optical data.  The observational results are
presented in \S3 starting with the key result of this work, namely the
multiwavelength spectrum of the quiescent black hole XTE~J1118+480.  A
remarkably similar spectrum of black hole A0620--00 is also shown.
In \S4 we present an ADAF model for the X--ray
spectrum and a thermal model for the optical/UV continuum.  The
suitability of these models is discussed in \S5.  In \S6 we offer our
conclusions, and we end by comparing the feeble luminosity of J1118 to
the luminosities of other quiescent X--ray novae with both black--hole
and neutron--star primaries.

\section{Observations and Analysis}

The times of the various {\it CXO, HST} and MMT observations are
summarized in Figure~1.  As shown in the figure, one set of
observations occurred during 2002 January 11, which we refer to below
as {\it epoch~1}.  The second set of observations were conducted on
the following day on January 12 (UT), which we refer to as {\it
epoch~2}.  As indicated by the dashed lines substantial simultaneous
coverage in the three wavebands was obtained during {\it epoch~2}.
The spectra reported herein are based solely on the totality of the
data obtained during {\it epoch~2} plus the {\it HST} FUV data obtained
during {\it epoch~1}.

\subsection{{\it Chandra} X--ray Observations}

\subsubsection{XTE~J1118+480}

The X--ray data were obtained with the Advanced CCD Imaging
Spectrometer (ACIS; Garmire et al. 1992) onboard {\it Chandra}.  Data
were analyzed using the {\it Chandra} X--ray Center (CXC) CIAO v1.1
software\footnote{ http://asc.harvard.edu/ciao/}.  The ACIS-S3
detector was operated in the standard configuration with a time
resolution of 3.24~s.  The data were filtered to include only pulse
heights from 0.3--7.0~keV in order to limit the background.  The
source counts were recorded at a position consistent with the optical
position of J1118.  Only counts that fell within a 1\fas5-radius
source extraction circle were selected, since 95\% of the source flux
is contained in such a circle for a point source observed on-axis (van
Speybroeck et al. 1997).  Only time intervals for which the background
rate (per $10^{5}$~pixels) was less than 0.35 counts~s$^{-1}$ were
selected for analysis, which resulted in a net exposure time of
45.8~ks for {\it epoch~2} and 7.7~ks for {\it epoch~1}.  A total of 80
source photons were detected in {\it epoch~2} and 9 source photons in
{\it epoch~1}. In the following we restrict our discussion to the {\it
epoch~2} observation which netted 80~source counts.  The predicted
number of background counts in the source extraction circle was small,
0.9 counts, and we neglected the background in our spectral analysis.

Source spectra were derived using HEASARC XSPEC v11.0\footnote{
http://heasarc.gsfc.nasa.gov/docs/xanadu/xspec/index.html} and also
SHERPA Version 2.1.2 within CIAO\footnote{
http://asc.harvard.edu/ciao/download/doc/sherpa\_html\_manual/index.html}.
The specific results reported herein are based on XSPEC, although the
two software packages gave consistent results.  Here we concentrate on
the analysis of the pulse-height data for {\it epoch~2}. The 80~source
counts were binned into 9 bins, each with a nominal 9~counts per bin,
in order to allow the use of $\chi^{2}$ statistics.  A response file
appropriate to the ACIS-S3 detector temperature ($-120$~C) was
used. Of special importance, the response file was corrected to the
date of the observations for the ongoing degradation in the ACIS-S low
energy quantum efficiency using the ``corrarf'' routine\footnote{
http://cxc.harvard.edu/cal/Acis/Cal\_prods/qeDeg/index.html}.  We
fitted the data using several single-component spectral models with
interstellar absorption (Balucinska-Church \& McCammon 1992).  We
fixed the column density to the value determined in outburst: $N_{\rm
H}~=~1.2~\times~10^{20}$~cm$^{-2}$ (McClintock et al. 2001b; Esin et
al. 2001).  The blackbody and disk blackbody models and the
Raymond-Smith model with cosmic abundances did not give acceptable
fits to the data; the values of reduced $\chi^{2}_{\rm \nu}$ ranged
from 2.0 to 2.9 for 7 dof. (However, if one allows $N_{\rm H}$ to vary
freely, then these models can be fit satisfactorily.)  The
bremsstrahlung model gave an acceptable fit with $kT$~=~1.70~keV
($\chi^{2}_{\rm \nu}$~=1.32 for 7 dof).  A slightly better fit was
obtained with a power--law model: photon index
$\Gamma~=~2.02~\pm~0.16$ ($\chi^{2}_{\rm \nu}$~=~1.26 for 7~dof).

We re-fit the power--law data allowing $N_{\rm H}$ to vary.  The
resultant values of both $N_{\rm H}$ and $\Gamma$ are within one
standard deviation of the values obtained above with $N_{\rm H}$
fixed; moreover, the improvement in $\chi^{2}_{\rm \nu}$ is negligible
(1.26 {\it vs.} 1.24).  Therefore, we adopt the power--law model with
the column density frozen at the value determined during outburst:
$N_{\rm H}~= 1.2~\times~10^{20}$~cm$^{-2}$.  For this model, the
absorbed energy flux is $8.7\times 10^{-15}$\flux, and the unabsorbed
flux is only 5\% larger (E~=~0.3--7.0~keV).  The unabsorbed luminosity
is $L_{\rm x}~\approx~3.5~\times~10^{30}$~\lum.  With $N_{\rm H}$
fixed, there is only one important parameter, the photon index
$\Gamma$.  We derived an X--ray error box in the $\nu$$F_{\rm \nu}$
{\it vs.} $\nu$ plane, which is presented in \S3, as follows (also see
Narayan et al. 1997a).  We computed a contour plot of the
normalization constant $K$ {\it vs.}  $\Gamma$ with a single contour
that encompasses the 90\% confidence level ($\chi^{2}_{\rm
total}~+~2.71$; Lampton, Margon \& Bowyer 1976).  We determined the
values of ($K,\Gamma$) at 60 points around this contour and computed
and plotted ($\nu$$F_{\rm \nu}$ {\it vs.} $\nu$) the corresponding
model spectra over the range 0.3--7~keV.  The X--ray error box is
defined by the outer envelope of this collection of spectra.

We fixed the absorbing column depth at the value determined during
outburst, $N_{\rm H}~= 1.2~\times~10^{20}$~cm$^{-2}$, for several
reasons. First, with only 80 detected X-ray photons and a UV source
spectrum that is not known {\it a priori}, we are unable to place
strong constraints on $N_{\rm H}$ from our observations.  As noted
above, however, our Chandra spectrum is entirely consistent with the
column depth determined in outburst.  Second, the interstellar column
was tightly constrained by spectral studies during outburst (Hynes et
al. 2000; McClintock et al. 2001; Frontera et al. 2001).  Our adopted
value of $N_{H}~=~1.2~\times~10^{20}$~cm$^{-2}$ is closely bracketed
by the most likely range for $N_{\rm H}$ of
$(1-1.5)~\times~10^{20}$~cm$^{-2}$ (e.g., Frontera et al. 2001).
Third, as we state in \S1 and the abstract, the absorption is small
and our conclusions are little affected by it.  For example, if one
adopts a simple power-law model and fixes $N_{\rm H}$ at the upper
limit of $1.5~\times~10^{20}$~cm$^{-2}$, the photon index does
increase somewhat from the value given above ($\Gamma~=~2.02$) to
$\Gamma~=~3.12~\pm~0.34$; however the latter fit is significantly
poorer, $\chi^{2}_{\rm \nu}$~=~1.90 ({\it vs.} $\chi^{2}_{\rm
\nu}$~=~1.26) for 7~dof. Similarly, the minimum transmission in the UV
for the wavelengths considered in the text is 82\% for $N_{\rm
H}~=~1.2~\times~10^{20}$~cm$^{-2}$ and it is negligibly different for
$N_{\rm H}~=~1.5~\times~10^{20}$~cm$^{-2}$.  Finally, despite the
relatively high inclination of the source, it is reasonable to assume
that absorption within the source is negligible in outburst and in
quiescence because in both cases the inner disk is truncated far from
the X-ray emitting plasma (McClintock et al. 2001; this work), and
because the hot accretion flow is almost certainly optically thin
(e.g. Esin et al. 2001; Markoff et al. 2001).  In any case, it is
difficult to imagine any plausible reason why the absorption in
quiescence should exceed the minuscule absorption that was observed in
outburst.

As noted above, we used $\chi^{2}$ statistics even though there were
only about 9 counts in each of the 9 pulse-height bins.  As a check on
our use of $\chi^{2}$ statistics, we refitted the {\it unbinned} data
with the power--law model and $N_{\rm H}$ fixed, as above, using
``C--statistics'' as implemented in XSPEC (Arnaud \& Dorman 2000, and
references therein).  The C--statistic is appropriate when there are
few counts per bin and the error on the counts is pure Poisson, as in
our case.  With this approach we found a very similar value of the
power--law index and an error that was only slightly larger than the
value quoted above: $\Gamma$(C--stat)~$=~1.92~\pm~0.17$.  As a second
check, we refitted the binned data using $\chi^{2}$ statistics with the
Gehrels weighting function (Gehrels 1986; Arnaud \& Dorman 2000).
This method is frequently used in the case that one has only several
counts per bin.  With this approach, we found that the value of the
photon index was unchanged, but the error was increased somewhat
(38\%): $\Gamma$(Gehrels)~$=~2.02~\pm~0.22$.  Thus the magnitude of
the error estimate is scarcely affected by the use of C--statistics and
increased only modestly by the use of Gehrels weighting.  In
conclusion, throughout this work we adopt the conventional approach to
spectral analysis discussed earlier in this section; namely, we use
$\chi^{2}$ statistics and a conventional $\sqrt(N)$ weighting function.

\subsubsection{A0620--00}

We extracted and reanalyzed a 44~ks data set for A0620--00 from the
{\it Chandra} archive.  The observation was made on 2000 February 29
using the ACIS-S3 detector.  The data modes and data analysis were
essentially identical to those described above for J1118, and the
results obtained are consistent with the results that have been
published for this data set (Garcia et al. 2001; Kong et al. 2002).
In brief, a 1\fas5-radius source extraction circle was used.  Only
time intervals for which the background rate (per $10^{5}$~pixels) was
less than 0.28~counts~s$^{-1}$ were selected for analysis.  A total of
119 source photons were detected during a net exposure time of
41.8~ks.  The predicted number of background counts in the source
aperture was 0.5, and we therefore neglected the background.  The
source counts were binned into 10 bins with a nominal 12 counts per
bin.  The response file was corrected as described above.  In fitting
the data, we fixed the column depth at $N_{\rm
H}~\approx~1.94~\times~10^{20}$~cm$^{-2}$, which corresponds to
$E_{\rm B-V}~=~0.35$~mag (Wu et al. 1983; Predehl \& Schmitt 1995). As
before, with $N_{\rm H}$ fixed there is only one important parameter,
the photon power--law index: $\Gamma~=2.26~\pm~0.18$.  We derived an
X--ray error box in the $\nu$$F_{\rm \nu}$~{\it vs.}~$\nu$ plane
precisely as described in \S2.1.1.

\subsection{{\it HST} Ultraviolet Observations}

The ultraviolet data were collected using the Space Telescope Imaging
Spectrograph (STIS) aboard {\it HST}.  The near--ultraviolet (NUV) data
were obtained as part of the simultaneous observations during {\it
epoch~2} (Fig. 1) using the G230L grating, the 52$''$~x~0\fas5
aperture, and the NUV-MAMA detector.  The spectral resolution of these
data is $\approx$~4.5~\AA\ (FWHM).  The far-ultraviolet (FUV) data
were obtained during {\it epoch~1} (Fig.~1) using the G140L grating,
the 52$''$~x~0\fas5 aperture, and the FUV-MAMA detector.  The spectral
resolution is $\approx$~1.5~\AA\ (FWHM).  All of the data were
recorded in ``time-tagged'' mode.  Both the NUV and the FUV
observations consisted of a series of five exposures obtained during
five consecutive (96--min) {\it HST} orbits.  The individual exposures were
nominally 2800~s in duration, and the total net exposure time for the
NUV and the FUV observations alike was 13.4~ks.  We were forced to
break our observing campaign into two epochs separated by one day
because a single observing session or ``visit'' using a MAMA detector
is limited to a maximum of five consecutive {\it HST} orbits.

The results presented herein are based on our analysis of the standard
data products produced by the STScI ``pipeline.''  However, because
both the NUV and FUV continuum fluxes are faint ($f_{\rm
\lambda}~\lesssim~2~\times~10^{-17}$\fluxlam) the ``pipeline'' process
failed to extract useful 1-D spectral files.  We therefore examined
the calibrated 2-D files, determined the Y-location of each trace, and
derived the spectra using the standard STSDAS analysis task within
IRAF\footnote{ IRAF (Image Reduction and Analysis Facility) is
distributed by the National Optical Astronomy Observatories, which are
operated by the Association of Universities for Research in Astronomy,
Inc., under contract with the National Science Foundation.} to extract
1-D spectra from 2-D images.  The source spectrum itself was extracted
with a box of standard width, 11 pixels.  The background regions on
each side of the source spectrum were extracted using a box of width
30 pixels and the buffer regions between the source and background
extraction boxes was 5 pixels.  The extracted spectra were found to be
fairly insensitive to these details.  The resultant five 1-D spectra
of each type (NUV and FUV) were averaged.  In Figure~2 this pair of
spectra are shown as a single composite spectrum.  The photometric
accuracy of the STIS/MAMA detectors is 0.04 mag; however, based on
several trial extractions of the 2-D data, we estimate that 0.10 mag
is a fairer estimate of the uncertainty at these low flux levels.

As is apparent in Figure~2, the FUV fluxes are a factor of a few times
greater than a simple extrapolation of the NUV continuum fluxes would
imply.  Since the photometric accuracy of the STIS/MAMA detectors is
very high (see above), we attribute this difference to source
variability that occurred between {\it epoch~1} and {\it
epoch~2}. This point is disucussed in \S3.1.

\subsubsection{NUV and optical emission lines}

The width of the Mg~II~2800~\AA\ emission line is
FWHM~$\approx~31$~\AA, which is much broader than the instrumental
resolution (4.5~\AA).  However, the Mg~II line is not the best
indicator of the Doppler line width because it is a doublet with a
spacing of 7.2~\AA.  The resolution of our broadband optical spectrum
(\S2.4.2) is too poor ($\sim~12$~\AA) to provide a useful measure of
the H$\alpha$ line width.  We therefore used a sum of eight spectra
obtained in 2000 December (McClintock et al. 2001a) with a resolution
of $\approx~3.6$~\AA\ (FWHM) to deduce the width of the H$\alpha$
line: FWHM~=~$53~\pm~2$~\AA~(2400~km~s$^{-1}$).  The full width at
zero intensity is $\sim~75$~\AA\ (3400~km~s$^{-1}$), and definitely
$<$~97~\AA\ (4400~km~s$^{-1}$).  The H$\alpha$ line has a clear
double-peaked structure that varies with orbital phase, as observed
for other quiescent black hole X--ray novae such as A0620--00 and Nova
Muscae~1991 (e.g., Orosz et al. 1994).  This double-peaked structure
is not apparent in the H$\alpha$ profile presented in \S2.4.2 because
of poor spectral resolution (12~\AA\ FWHM) or in the Mg~II line in
Figure~2 because it is a doublet.  We conclude that the broad lines of
Mg~II and H$\alpha$ and the double-peaked profile of H$\alpha$ differ
in no significant way from similar lines observed in cataclysmic
variables and other quiescent X--ray novae, where they are attributed
to emission from a Keplerian accretion disk (Horne~\& Marsh 1986;
Orosz et al. 1994; Marsh, Robinson \& Wood 1994).

\subsection{FUV emission lines}

In addition to the Balmer and Mg~II lines, two additional emission
lines were detected in the FUV band: N~V~1240~\AA\ and Si~IV~1403~\AA.
These lines are among the commonly observed UV emission lines
(e.g., Teays 1999) and they are present in spectra of J1118 obtained
during outburst (Haswell et al. 2002).  The Si~IV line can be seen as
a weak feature in the FUV spectrum in Figure~2.  The N~V line does not
appear in the same spectrum because we were unable to reliably extract
the part of the continuum spectrum sandwiched between Ly$\alpha$ and
O~I~1302~\AA, which are two very intense geocoronal lines.  To assess
the statistical significance of the N~V and Si~IV lines, we first
averaged the five 2-D FUV spectral images (i.e., the flat-fielded
science images) and used SAOImage/DS9 to measure the counts along the
spectrum and in the adjacent background.  The net source counts {\it
vs.} pixel number (or wavelength) is shown in Figure~3 for the two
lines.  Both the N~V and the Si~IV lines are present at a $4.5~\sigma$
level of confidence.  Using the extracted spectra binned at 2~\AA, we
measured the wavelength of N~V to be 1241~\AA\ and the wavelength of
Si~IV to be 1399~\AA\ with estimated errors of $\pm~4$~\AA.  These
values are in reasonable agreement with the established wavelengths of
these lines, which are mentioned above.

\subsubsection{Ultraviolet orbital light curves}

We have five individual NUV spectra of J1118 that span about two
4.1-hr orbital cycles of the source.  Similarly we have five NUV
spectra of A0620--00 which span one complete 7.8-hr orbital cycle
(McClintock \& Remillard 2000).  For each spectrum of J1118 we
determined the intensity of the continuum.  The result is shown in
Figure~4a, where the data are plotted twice (solid dots) as a function
of photometric phase.  We made precisely the same measurement on the
five spectra of A0620--00 and the results are plotted in Figure 4b.
Similar measurements were made of the Mg~II line intensity for both
sources, and these results are plotted as open circles in Figures 4ab.
The general appearance of the Mg~II and continuum light curves is
similar; however, the uncertainties and the apparent scatter in the
Mg~II data are larger, and we therefore focus on the continuum light
curves.  (Additional measurements of the FUV continuum intensity and
the H$\alpha$ line intensity for J1118 proved too noisy to be useful.)
The amplitude of the continuum light curve is large:
$\approx~0.35$~mag for J1118 and $\approx~0.25$~mag for A0620--00.

Note that the intensity varies smoothly with orbital phase for both
J1118 and A0620--00 (Figs. 4ab).  However, if the same intensity data
for J1118 (which spans two orbital cycles) is plotted as a function of
time, the light curve is erratic, as shown in Figure~5.  This result
suggests that the variations are indeed tied to the orbital cycle.
(This test is meaningless for A0620--00 since these data span only one
orbital cycle.)

\subsection{MMT Optical Spectroscopy}

All of the spectroscopic data were obtained using the 6.5 m MMT and
the Blue Channel Spectrograph equipped with a Loral CCD (3072~x~1024)
detector.  The bulk of the observations were made during both {\it
epoch~1} and {\it epoch~2} (Fig. 1); some additional data were
obtained a night later on January 13 (09:50 -- 13:30 UT).  Important
intermediate objectives included a determination of the orbital phase
of J1118 and the ``rest-frame'' spectrum of the secondary star.  (We
also make use of the orbital phase in \S2.3.1 and Figure~4a to interpret the UV
light curve of J1118.)  Our primary objective was to derive the
optical spectrum of the accretion flow surrounding the black hole
(i.e., the spectrum of the accretion disk and/or the ADAF).  These
objectives required the use of two instrumental setups.  We first
obtained medium-resolution spectra (1.4~\AA\ FWHM; 5000--6400~\AA) of
J1118 and 16 spectral comparison stars in order to determine the
spectral type of J1118, and then we obtained a broadband
(3400--6800~\AA), low-resolution ($\approx~12~$\AA\ FWHM) spectrum of
J1118.  Using these two data sets we decomposed the broadband spectrum
into a stellar spectrum and a residual (disk/ADAF) spectrum following
the method described by Marsh, Robinson, \& Wood (1994).

\subsubsection{Medium-resolution spectrum}

All of the medium-resolution spectra were obtained with a 1\fas0 slit
and a 1200 groove~mm$^{-1}$ grating in first order. On 2002 January 13
UT we obtained ten such spectra of J1118 interspersed with exposures
to a calibration lamp (He--Ne--Ar) and additional observations of flux
standard stars.  On this night, the seeing was $\approx$~1$''$ and the
sky was clear.  The standard spectral reductions were done using IRAF.
The individual source exposure times were 900~s, and the observations
of J1118 extended from 9:50~UT to 12:45~UT, which corresponds to about
70\% of an orbital cycle of the source.  We used the spectrum of the
K5V velocity standard GJ563.1 (see below) as a cross-correlation
template to derive a radial velocity curve for the secondary star
(e.g., McClintock et al. 2001a), which is shown in Figure~6.  We
adopted an orbital period of $P$~=~0.16994~days (Wagner et al. 2001;
McClintock et al. 2001a) and fitted a sine function to the velocity
data, thereby determining the heliocentric time of maximum redshifted
velocity of the secondary to be $T_{\rm
0}$~=~HJD~$2,452,287.9929~\pm~0.0005$~d.  Using these orbital
elements, the ten spectra were Doppler shifted to zero velocity and
combined to yield the rest-frame spectrum shown in Figure~7a.  Our
value of $T_{\rm 0}$ is also crucial to the
interpretation of the NUV light curve of J1118 (\S2.3.1).

We also obtained 20 spectra of 16 bright, spectral-comparison or
template stars at 1.4~\AA\ resolution (FWHM) during {\it epoch~1} when
the seeing was poor and after J1118 had set on the nights of
January~12 and January~13 UT.  The template stars were carefully
chosen to have well-determined spectral types and metallicities.  The
spectra were obtained with the 1\fas0 slit and the 1200
groove~mm$^{-1}$ grating, and each observation was paired with an
exposure of the wavelength calibration lamp.  The template spectra and
the rest-frame spectrum were all Doppler shifted to the same relative
velocity.  The spectrum of one of these comparison stars, GJ563.1, is
shown in Figure~7c.

\subsubsection{Broadband spectrum}

This component of the MMT observations was scheduled to be
simultaneous with the space-based observations (Fig.~1).  During the
{\it epoch~1} observations of J1118, the seeing was poor
($\gtrsim5''$); the broadband data we collected were not useful and
we disregard them.  All of the useful broadband, spectrophotometric
observations were made during {\it epoch~2} with a 5\fas0 slit and a
300 groove~mm$^{-1}$ grating in 1st order; the seeing ranged from
$\approx2.5-4''$ throughout the night and the sky was clear.  A
total of 13 useful 15-min broadband spectra of J1118 were obtained
between 06:25 and 12:40 UT; they span 1.5 orbital cycles of J1118.
Observations were interspersed with exposures to a wavelength
calibration lamp (He--Ne--Ar) and three observations of a flux standard
star, Feige~34 (Oke 1990).  The standard spectral reductions were
performed using the software package IRAF.  The 13 spectra were
averaged and the resultant spectrum is shown in Figure~8a.

\subsubsection{Resultant spectrum of the accretion disk/ADAF}

We used the 20 template spectra mentioned above and followed the
method outlined in Marsh, Robinson, \& Wood (1994) to decompose the
rest-frame spectrum into its disk and stellar components.  The
dereddened rest-frame spectrum and the template spectra were all
normalized to unity at 5500~\AA.  The template spectra were then
scaled and subtracted from the rest-frame spectrum.  An rms difference
was computed for each template spectrum, in precisely the manner
described in Orosz et al. (2002).  As shown in Figure~9, this rms
difference decreases monotonically from 0.068 at spectral type M3 to
0.054 at K5, and it changes very little between K7 and K5.  In this
way, we find that the secondary has a spectral type near K5, appears
to have a somewhat low metallicity ([Fe/H]~=~$-1$), and contributes
$45~\pm~10$\% of the total light at 5500~\AA.  Unfortunately, we do
not have spectra of any template stars earlier than K5.  However
spectral types earlier than K5 are disfavored by other observers
(McClintock et al. 2001a; Wagner et al. 2001).  Thus we adopt the
somewhat metal poor K5 dwarf GJ563.1 as a proxy for the secondary
star; its spectrum is shown in Figure~7c.  As expected, the difference
between the spectrum of GJ563.1 and the spectrum of J1118, which is
shown in Figure~7b, is fairly featureless, except in the vicinity of
the Mg~b complex ($\sim5175$~\AA). There the lines in J1118 are
somewhat stronger than the lines in the proxy star.  The other major
feature in the difference spectrum is broad emission at He I 5875~\AA,
which is produced in the accretion disk of J1118.

We note that a K5/7V star ($R_{2}~\approx0.7$\Rsun~) would not fit in
its Roche lobe for a 4.1-hr orbital period ($R_{\rm
L}\approx0.5$\Rsun~).  However, this is not a problem for several
reasons.  First, a metal-poor subdwarf is significantly smaller than a
ZAMS star of the same spectral type.  For example, our proxy star
GJ563.1 has a radius that is $<$80\% of a K5V star, based on its
Hipparcos parallax (Perryman et al. 1997).  Second, the companion star
has had a tumultuous history and is not a pristine ZAMS star (e.g., de
Kool et al. 1987).  The large mass ratio observed for J1118,
$Q~=~27~\pm~5$ (Orosz 2001), can be used to illustrate this point:
Assuming a minimum mass ratio of $Q~=~22$ and a generous primary mass
of $M_{1}~=8$~\Msun~, one finds $M_{2}~<~0.36$~\Msun~.  For a ZAMS
star this mass would imply a spectral type later than M2, which is
definitely ruled out (Fig.~9; McClintock et al. 2001; Wagner et
al. 2001).  Finally, in the context of the secondary's evolutionary
history, it is not surprising that the star has a relatively early
spectral type.  Following de Kool et al., we suppose that the inital
mass of the secondary was perhaps double its present mass.  During the
later stages of its evolution the star loses much of this mass from
its envelope and its orbit shrinks.  Since the energy generation in
the core of the star is little affected by the mass loss from its
envelope, the star will have about the same luminosity it started
with.  However, since its radius is smaller, it will be somewhat
hotter. In short, its spectal type will be somewhat earlier than it
was initially despite the fact that its mass is now much less.

The success in deconvolving the spectral components in the
medium-resolution spectrum allows us to derive the broadband spectrum
of the disk/ADAF as follows.  We subtracted a Kurucz model of a K5
dwarf with [Fe/H]~=~$-1$ from the broadband spectrum discussed above
(Fig.~8a).  The resultant spectrum of the non--stellar or residual
component of emission is shown in Figure~8b.  At the blue end, the
flux is almost entirely due to this residual component and the
estimated uncertainty in the flux there is 0.1~mag, as quoted above.
However, at longer wavelengths the residual spectrum depends more
strongly on the model spectrum and the uncertainties are larger.
Given our reliance on a synthetic model spectrum and our lack of
template spectra earlier than K5, we estimate that the fluxes are
uncertain by as much as 0.3 mag at the extreme red end of the residual
spectrum (Fig. 8b), with the uncertainty increasing roughly linearly
with wavelength from 0.1 mag at the blue end.

\section{Multiwavelength Spectra}

\subsection{XTE~J1118+480}

The complete multiwavelength spectrum of J1118 is shown in Figure 10.
All of the data were obtained essentially simultaneously during {\it
epoch~2} except for the FUV data, which were obtained one day earlier
(Fig. 1).  Very modest corrections for reddening and absorption have
been applied (the maximum UV and X--ray interstellar attenuations are
18\% and 30\%, respectively; see \S1), and the spectrum is presented
in units of log($\nu$$F_{\rm \nu})$ {\it vs.} log$(\nu)$, which have
been used extensively in modeling the spectra of X--ray novae (e.g.,
Narayan et al. 1997a).

We now discuss in turn the three spectral components -- X--ray,
ultraviolet and optical.  The best-fitting X--ray model is represented
by a heavy horizontal line which corresponds to a pure power--law
spectrum with photon index $\Gamma~=~2.02~\pm~0.16$ (\S2.1.1). The
bowtie-shaped X--ray error box is drawn at the 90\% level of
confidence; its derivation is discussed in \S2.1.1.  The unabsorbed
0.3--7~keV~luminosity is $L_{\rm x}~=~3.5~\times~10^{30}$\lum~or
$4~\times~10^{-9}L_{\rm Edd}$ for $D~=~1.8$~kpc and $M~=7M_{\odot}$
(McClintock et al. 2001a).

The {\it HST} ultraviolet spectrum (log($\nu)~\gtrsim~15.0$) appears in
two segments, which correspond to the two spectra (NUV and FUV) shown
in Figure~2. The NUV spectrum, which can be easily identified by its
prominent Mg II 2800~\AA\ line, has been boxcar-smoothed to 20~\AA\
(FWHM), and the FUV spectrum, which plunges steeply downward to
log($\nu$$F_{\rm \nu}) \sim -~14.2$, has been smoothed to 10~\AA\
(FWHM).  Data with especially low signal-to-noise have been trimmed
from the extremities of the NUV spectrum (cf. Fig. 2). In assembling
the composite spectrum shown in Figure 10, the following important
normalization correction has been made to the FUV data, which were
obtained a day before all of the other data (Fig. 1).  Since the X--ray
intensities of quiescent X--ray novae are known to vary by up to an
order-of-magnitude on a one-day time scale (Wagner et al. 1994; Garcia
et al. 2001), we have taken the liberty of dividing the FUV fluxes by
the factor 2.63 to correct for the probable effects of source
variability.  This factor was determined by making linear fits to both
the NUV and FUV continuum spectra [log($\nu$$F_{\rm \nu}$) {\it vs.}
log($\nu$)], and by matching these fits at log(${\nu}$)~=~15.2.  Based
on the fits, we note that the FUV spectrum is slightly steeper
(slope~=~2.77) than the NUV spectrum (slope~=~2.59).

The MMT optical spectrum of the disk/ADAF component smoothed to
20~\AA\ (FWHM), which is shown in Figure 8b, appears at the lowest
frequencies.  It was derived by subtracting a K5V model spectrum from
the MMT spectrum shown in Figure~8a (\S2.4.3.).  The most prominent
line is of course H$\alpha$; some higher members of the Balmer series
are also evident (cf. Fig.~8).  Despite the limitations of the optical
data, which are discussed in \S2.3, it is reassuring that they match
up satisfactorily with the NUV data in Figure~10 without adjusting the
normalization of either data set.

The spectrum in Figure~10, which is the central result of this paper,
is comprised of two components: (1) a power--law component, which we
model and discuss in \S4.1 \& \S5.1, and (2) an optical/UV component,
which we model and discuss as a thermal, accretion-disk component in
\S4.2 and \S5.2.  Before turning to these subjects, we complete this
section by presenting a multiwavelength spectrum of black hole
A0620--00 (\S3.2). 

\subsection{Comparison of XTE~J1118+480 and A0620--00}

Limited multiwavelength data of high quality exist for only one other
quiescent X--ray nova -- the short--period system A0620--00 ($P_{\rm
orb}~=~7.8$~hr).  In Figure~11 we compare these data, which were not
obtained simultaneously, to the spectrum of J1118.  An {\it HST}
STIS/NUV spectrum of A0620--00 is shown with its prominent Mg~II
2800~\AA\ line.  These data, which were published previously
(McClintock \& Remillard 2000), are binned here in 20~\AA\ bins to
facilitate their comparison with the spectrum of J1118.  The best-fit
X--ray spectrum of A0620--00 is represented by the heavy line, which
corresponds to the power--law model discussed in \S2.1.2.  The
adjacent curved lines bound the 90\% confidence error box.

There are striking similarities between the spectra of the two
sources.  The NUV spectra are almost indistinguishable in slope and
spectral content, except that the Fe~II emission feature
(2586--2631~\AA) located just to the right of the dominant Mg~II line
is stronger in A0620--00 (McClintock \& Remillard 2000) than in J1118.
Turning to the X--ray, the spectral indices of the two sources appear
to differ slightly: $\Gamma~=2.02~\pm~0.16$ for J1118 and
$\Gamma~=2.26~\pm~0.18$ for A0620--00.  In fact, however, the spectral
index of A0620--00 is more uncertain than the error just quoted implies
since it also depends on the value of the reddening.  For example, if
the value of the reddening that we adopted ($E_{\rm B-V}$~=~0.35~mag)
were to be decreased by 30\%, the best-fit photon index would exactly
match the value quoted for J1118 (i.e., $\Gamma~=~2.02$).  Thus, given
the statistical uncertainties and the uncertainty in the reddening of
A0620--00 ($\approx~0.05$~mag), the two X--ray spectra are effectively
indistinguishable.  Finally, the relative intensities of the two
sources in the two bands are very comparable, especially given that
the A0620--00 observations were not simultaneous.  A0620--00 is about 6
times brighter than J1118 in the UV and about 4 times brighter in the
X--ray.  One expects A0620--00 to be about a factor of 3 brighter than
J1118 given its smaller estimated distance: $\sim~1.0$~kpc {\it vs.}
$\sim~1.8$~kpc, respectively.

{\it The remarkable similarities between the spectra of the two
sources argue strongly that our spectrum of J1118 represents the
canonical spectrum of a stellar--mass black hole radiating at $L_{\rm
x}~\sim~10^{-8.5}L_{\rm Edd}$.}

\section{Models}

\subsection{ADAF Model of the X--ray Spectrum}

Narayan et al. (1996) showed that an accretion model consisting of a
thin accretion disk at large radii and an ADAF at small radii, with a
low mass accretion rate $\dot M$, has an X--ray spectrum that
resembles the observed spectrum of the binary black hole system
A0620--00 in quiescence.  Subsequently, Narayan et al. (1997a) and Hameury et
al. (1997) found that a similar model also explains the X--ray spectra
of two other black hole systems, V404 Cyg and GRO J1655--40.  In view
of these earlier successes, it is of interest to test the model
against the data we have obtained on XTE J1118+480.

Since the time of the above papers, there have been some improvements
to the ADAF model.  First, following the work of Nakamura et
al. (1997), it is now customary to model self-consistently the
advection of energy by electrons (Esin et al. 1997); prior to this
work only energy advection by ions was considered.  Second, following
the work of Stone, Pringle \& Begelman (1999) and Igumenshchev \&
Abramowicz (2000), it is now recognized that ADAFs do not necessarily
accrete all the mass supplied to them.  Significant mass could be lost
in a strong outflow (Narayan \& Yi 1994, 1995a; Blandford \& Begelman
1999), or the accretion could be suppressed by convection (Narayan,
Igumenshchev \& Abramowicz 2000; Quataert \& Gruzinov 2000;
Igumenshchev, Abramowicz \& Narayan 2000).  Both effects may be
modeled by using a standard ADAF model but writing the mass accretion
rate as the following function of radius (Quataert \& Narayan 1999):
$$ 
\dot M(r) = \dot M(r_{\rm tr}) \left( {r\over r_{\rm tr}} \right)^p, 
\qquad r \le r_{\rm tr}, \eqno (1)
$$ 
where $p$ is a free parameter.  Here, $r_{\rm tr}$ is the transition
radius between the outer thin disk and the ADAF, and $\dot M(r_{\rm tr})$
is the rate at which mass is supplied to the ADAF at $r_{\rm tr}$,
presumably by disk evaporation.  The ADAF models used in the early
work on quiescent X--ray novae corresponded to $p=0$, but we consider
here models over the range $0 \le p \le 1$.  Here and elsewhere, the
dimensionless radius $r$ is related to the physical radius R by
$r~=~R/R_{\rm S}$, where $R_{\rm S}~=~2GM/c^{2}$ is the Schwarzschild
radius. 

Quataert \& Narayan (1999) studied models with non-zero $p$ and showed
that the spectra of these models are approximately degenerate with
respect to two model parameters.  One of the parameters is $p$
introduced above, and the other is $\delta$, the fraction of the heat
energy released by viscous dissipation that goes into electrons in the
accreting plasma (the remainder, $1-\delta$, goes into the ions).
Quataert \& Narayan (1999) found that spectra remain nearly invariant
along diagonals in the $p-\delta$ plane.  Motivated by this study, we
have computed a grid of spectral models of XTE J1118+480 for different
choices of $p$ and $\delta$.

We take the mass of the black hole to be $7M_\odot$ and the
inclination of the system to be 80$^{\rm o}$.  As described in
\S2.2.1, the full-width at zero intensity of the H$\alpha$ line was
$\sim 3400 ~{\rm km\,s^{-1}}$ and definitely $< 4400 ~{\rm
km\,s^{-1}}$ in early 2000 December.  Assuming a Keplerian model for
the line-emitting gas, the two velocities correspond to radii of
$15000 R_{\rm S}$ and $8800 R_{\rm S}$, respectively. It is clear that
the transition radius between the ADAF and the disk cannot be larger
than these radii.  Therefore, in the models we set $R_{\rm tr}=10^4
R_{\rm S}$.  This is a ``minimal disk'', which contributes modestly
only in the infrared (Narayan et al. 1997a).  Our estimate of the
transition radius is based on observations of the H$\alpha$ line made
about one year prior to the observations reported here.  During this
period, J1118 declined further (Zurita et al. 2002) and it is likely
that the transition radius moved outward somewhat (Esin, McClintock,
\& Narayan 1997).  This, however, does not affect our models:
Increasing the radius has a negligible effect on the ADAF spectrum at
all frequencies (Narayan et al. 1997a) and the emission from the disk
becomes even more negligible.  We consider other choices for $r_{\rm
tr}$ in the next subsection.  We further assume that the mass
accretion rate in the outer disk is equal to $\dot M(r_{\rm tr})$, and
is independent of $r$.  The assumption of a constant $\dot M$ in the
disk is incorrect for a quiescent disk (e.g., Hameury et al. 1998),
but this assumption has a negligible effect on the computed spectrum
because the disk emission is very weak in the models.

The modeling techniques we have used are identical to those employed
in Quataert \& Narayan (1999).  For each choice of $p$ and $\delta$,
we adjust $\dot M(r_{\rm tr})$ until the predicted X--ray flux agrees
with the observed flux.  We then compare the shape of the calculated
X--ray spectrum with the observed spectrum to decide whether or not a
particular model is acceptable.  We have computed models over the
entire likely range of $p$ values, from 0 to 1, but we restricted
$\delta$ to the range 0 to 0.7 (we are not confident of our modeling
techniques for larger values of $\delta$).  Figure 12 shows a summary
of the results, with the solid dots indicating acceptable models and
the crosses indicating models that are ruled out.  Note that the
acceptable models tend to lie along a roughly diagonal line in the
$p-\delta$ plane, in agreement with the pattern found by Quataert \&
Narayan (1999).  Since there are several approximations in the
modeling, the precise location of the band of acceptable models is
hard to determine reliably.  However, we believe that the general
pattern is robust.

Figure 13 shows spectra corresponding to the models that we consider
to be consistent with the X--ray data (solid dots in Fig. 12).  All the
model spectra agree with the data in the sense that they lie entirely
within the X--ray error box.  Notice, however, that none of the models
gives a pure power--law spectrum in the X--ray band.  We discuss this
further in \S5.1.  Figure 14 shows representative examples of models
that do not fit the X--ray spectrum.  These models predict spectra that
are either too hard or too soft in the X--ray band.

\subsection{Thermal Model of the Optical Continuum}

The models described in the previous subsection have their emission
dominated by the ADAF.  The contribution from the disk is quite weak
because of the large transition radius we assumed.  In the optical/UV
band, the computed emission is predominately synchrotron radiation
from the relativistic electrons in the ADAF.

In the literature, there have been conflicting opinions on the
importance of the ADAF for the optical/UV emission.  In the original
paper by Narayan et al. (1996), the ADAF was found to be relatively
weak in the optical, and the emission came mostly from the outer disk.
However, in the later work of Narayan et al. (1997a), it appeared that
the ADAF alone could explain both the X--ray and optical/UV emission.
Among the models shown in Figure~13, we see both kinds of models; the
models with lower values of $p$ predict substantial ADAF emission in
the optical/UV, leaving little room for disk emission, whereas the
models with larger values of $p$ require substantial disk emission to
explain the optical/UV flux.

The presence in the optical/UV spectrum of strong emission lines,
which have to be produced by cool gas in the disk, suggests that much
of the optical/UV continuum may indeed be from the disk.  One way to
arrange this is to move the transition radius between the disk and the
ADAF to smaller radii.  This will enhance the importance of the disk
emission without having any significant effect on the ADAF.  Since the
emission from the ADAF is unaffected, we ignore it here for simplicity
and model the disk emission via the multi--temperature disk--blackbody
model spectrum of a steady accretion disk as implemented in XSPEC
(Arnaud \& Dorman 2000; Mitsuda et al. 1984; Makishima et al. 1986).
The model is specified by two parameters: (1) the temperature at the
inner edge of the disk, $T_{\rm in}$, and (2) a normalization
constant, $K$.

Three models are shown superimposed on an expanded plot of the
optical/UV data in Figure~15a.  The models are meant to be illustrative;
they are not fits to the data.  We have adjusted the normalizations of
the models to match the observed flux at log($\nu$)~=~15.05, where the
observations are judged to be most secure.  The model for $kT_{\rm
in}$~=~1.1~eV or $T_{\rm in}$~=~12800~K appears to best represent the
data, whereas the models for $kT_{\rm in}$~=~0.9~eV (10,400~K) and
$kT_{\rm in}$~=1.3~eV (15,100~K) appear to be poorer representations
of the data.  We note that if we were to ignore the FUV data, which
requires an {\it ad hoc} normalization (\S3.1), the inferred
temperature would be only slightly lower ($kT_{\rm
in}~\approx~1.0$~eV; Fig.~15a).  For our favored model with $T_{\rm
in}$~=~12,800~K, the value of the normalization constant is
$K~=~5.4~\times~10^{9}$.  This constant determines the inner disk
radius to be $R_{\rm in}~=~3.2~\times~10^{9}~$cm~$=~1500R_{\rm S}$,
where we have adopted $D~=~1.8$~kpc, $i~=80^{\rm o}$ and $M_{\rm
x}~=~7M_{\odot}$ (McClintock et al. 2001a).  The bolometric luminosity
of this model is $L_{\rm opt/UV}~=~6.5~\times~10^{31}$~\lum.

Another possibility is that the optical/UV continuum (and line
emission) is produced in some locally hot region of the disk, such as
the bright spot where the mass transfer stream from the secondary
strikes the disk.  In this case, the emission might be closer to a
single-temperature blackbody, and so we have compared the data for XTE
J1118+480 with such a model.  As shown in Figure~15b, the agreement is
significantly worse than for a disk blackbody spectrum, suggesting
that, if the emission comes from a localized region, the gas must be
multi-temperature.

\section{Discussion}

The multiwavelength spectrum of J1118 (Fig. 10) is robust because it is
immune to the effects of source variability (apart from the
non-simultaneous FUV observations).  Moreover, it is essentially
unaffected by absorption in the ISM since the transmission of the ISM
is $\geq~70$\% at all wavelengths.  This spectrum of J1118 (Fig. 10) is
comprised of two components: a hard X--ray part with an unabsorbed
0.3--7~keV luminosity of $L_{\rm
x}~\approx~3.5~\times~10^{30}$~\lum~and a soft optical/UV part with a
bolometric luminosity about 20 times larger that is punctuated by
several strong emission lines.  A key question, which we now discuss,
is the origin of these two emission components and their relationship.

\subsection{X--ray Component}

In our view, there can be little doubt that the X--ray emission in
quiescent black hole binaries comes from hot electrons near the black
hole.  Bildsten \& Rutledge (2000) suggested otherwise, proposing that
most of the X--rays may be emitted by a corona in the secondary.  This
proposal has been ruled out by measurements of both the luminosities
and spectra of several quiescent black hole systems (Narayan et
al. 2002; Garcia et al. 2001; Kong et al. 2002; Hameury et al. 2002).
A related proposal by Nayakshin \& Svensson (2001), that the emission
comes from a corona above the outer regions of the disk, faces
similarly difficulties, and is moreover ruled out by the presence of
eclipses in the X--ray light curves of some dwarf novae (see Garcia et
al. 2001, Hameury et al. 2002, for details).

Even after one has decided that the emission is from electrons near
the black hole, there are still choices to be made in developing a
model for the radiating gas.  The two major questions are the
following: (1) Does the emission come from the accretion flow or is it
from an outflow/jet?  (2) Do the emitting electrons have a thermal or
a nonthermal (e.g., a power--law) distribution of energies?

The ADAF model described in \S4.1 assumes that the emission is from
the accretion flow and that the electrons are thermal.  It is in some
sense a minimal model, with the great virtue that the flow properties
of the gas and the emission can be calculated fairly robustly.  In
particular, the thermal assumption strongly constrains the model,
since the radiation properties of a given gas element are completely
prescribed by a single number, the local temperature, which itself can
be calculated from an energy equation for the electrons.

Unfortunately, even the thermal ADAF model has two important but
undetermined parameters, $p$ and $\delta$.  Applying the model to a
given source, therefore, requires studying a grid of models in these
two parameters (Fig. 12).  In principle, numerical simulations should
be able to determine $p$ (see Stone et al. 1999, Igumenshchev \&
Abramowicz 2000, Igumenshchev et al. 2000, 2003, Stone \& Pringle
2001, Machida et al. 2000, 2001, Hawley, Balbus \& Stone 2001, for
examples of recent work), and detailed analyses of particle heating
such as the study of Quataert \& Gruzinov (1999, and references
therein) might provide a reliable estimate of $\delta$.

The good news for the thermal ADAF model is that it is consistent with
the X--ray spectrum of XTE J1118+480 over a comfortable range of the
$p-\delta$ parameter space (Fig.~12).  It is reassuring that such a
simple model is able to explain the observations fairly well.  Note,
however, that the predicted spectra of all the acceptable models are
curved (Fig. 13), and none of the models predicts the canonical
power--law spectrum that one is familiar with for more powerful X--ray
binaries.  If future observations with better signal-to-noise could
measure the sense of the curvature and its magnitude, they would allow
a tighter determination of $p$ and $\delta$.  Alternatively, if the
data were to indicate that the spectrum is of power--law form in the
X--ray band, it would severely compromise the thermal ADAF model as
described here.

The curvature in the predicted spectra of the ADAF model is primarily
due to the assumption of a thermal energy distribution for the
electrons.  This, coupled with the high electron temperature, causes
the spectrum to consist of a sequence of distinct Compton peaks
separated by valleys.  In this context note that, even though the
quiescent state and the low state of black hole binaries are closely
related according to the ADAF model (Esin et al. 1997, 1998), in that
both have a similar geometry for the flow and involve thermal
electrons, the predicted spectra are quite different.  In the low
state, the electron temperature is lower and the electron density is
substantially higher (because of the larger $\dot M$), and
consequently the multiple Compton peaks overlap with each other so as
to give a very accurate power--law spectrum (e.g., Sunyaev \& Titarchuk
1980).  Indeed, in this regime the spectrum depends only on a single
parameter, the Compton $y$ parameter, rather than on the temperature
and the density individually.  In the quiescent state, on the other
hand, the temperature is so high that the different Compton peaks are
distinct.  This allows more opportunity for testing models with
careful observations.  In particular, one could hope to measure the
temperature and the density separately, and also test the thermal
assumption.  Unfortunately, the present data are not accurate enough
for such detailed comparisons.

How realistic is the assumption of a thermal energy distribution for
the electrons?  It is hard to answer this question since the exact manner in
which the electrons are heated by viscous dissipation is not
understood (Quataert \& Gruzinov 1999).  Indeed, it is easy to
visualize heating mechanisms that would heat electrons efficiently to
produce a nonthermal energy distribution (Bisnovatyi-Kogan \& Lovelace
1997).  Mahadevan \& Quataert (1997) showed that Coulomb coupling
between electrons and ions as well as self-absorbed synchrotron
emission can cause nonthermal electrons to relax to a thermal
distribution.  However, the thermalization works only at relatively
high electron densities.  For the very low densities that we have in
our models, especially the models with larger values of $p$ and
$\delta$, the mechanisms considered by Mahadevan \& Quataert (1997)
are not likely to be efficient.  This suggests that the accreting gas
in an ADAF may contain nonthermal electrons.  It would thus be of
considerable interest to study ADAF models with nonthemal electron
distributions.  Such models would have more degrees of freedom than
the minimalistic thermal ADAF model we have considered, but they are
likely to predict quite different spectra (e.g., power--law) in the
X--ray band, and would therefore be useful for comparison with future
observations.

We note in this context a recent model of Sgr A$^*$ that invokes a
hybrid distribution of electrons consisting of both thermal and
nonthermal electrons (Yuan, Quataert \& Narayan 2003). It is likely
that models of this kind would be able to fit the quiescent spectrum
of J1118.  From a nonthermal ADAF model, it is a short step to a
different class of models, namely the jet model (e.g., Falcke \&
Biermann 1995), which again has more degrees of freedom than the
thermal ADAF model and which often (though not always) invokes a
power--law distribution of electrons.  The main difference is that in
the jet model the radiating electrons are located in a relativistic
outflow rather than in the accretion disk.  (The models assume that
the accretion flow is even more advection--dominated and dimmer than
the standard ADAF model).  Markoff, Falcke \& Fender (2001) have
developed a successful jet model for XTE J1118+480 during outburst
(when the source was in the low state).  With some changes in
parameters, the same model could presumably be applied to the
quiescent data presented here.  Being a nonthermal model, we expect it
to predict a power--law spectrum in the X--ray band, just as with any
ADAF model involving nonthermal electrons.  Indeed, we suspect that it
will be hard to distinguish between the two models.  For instance, the
ADAF model of Sgr A$^*$ mentioned above (Yuan et al. 2003) is quite
similar in its predictions to some jet models (Markoff et al. 2001).

It should be pointed out that the parameters of the jet model vary
considerably from one application to another.  For example, the model
makes use of mono-energetic electrons for the quiescent emission from
the Galactic Center black hole Sgr A$^*$ (Yuan et al. 2002), whereas
it invokes power--law electrons for most other applications.
Moreover, the power--law index $s$ of the electrons varies widely from
one application to the next: $s=2.0-2.6$ for the shock-flare model of
the X--ray flare in Sgr A$^*$ (Markoff et al. 2001), $s=1.5-2.0$ for
XTE J1118+480 in the low state (Markoff, Falcke \& Fender 2001), and
$s=2.8$ for NGC 4258 (Yuan et al. 2002).

\subsection{Optical/UV Component}

Three lines of evidence indicate that at least a substantial portion
of the optical/UV emission from J1118 
originates in an accretion disk and/or disk
structure (e.g., the bright spot): (1) The presence of the broad,
intense Balmer and Mg~II emission lines (Figs. 2, 8 \& 15); (2) the
$\approx0.35~$mag modulation of the NUV continuum (Fig. 4a); and (3)
the shape of the continuum spectrum (Fig.~15).  We discuss these
points in turn.

First, the great breadth of the H$\alpha$ and Mg~II lines and the
double-peaked structure of H$\alpha$ (\S2.2.1) argue forcefully
that these lines are created in a Keplerian accretion disk (Horne \&
Marsh 1986).  We accept the presence of these intense, broad lines as
rather direct evidence that an accretion disk is present in J1118,
A0620--00 and other quiescent X--ray novae (e.g., Orosz et al. 1994;
Marsh, Robinson \& Wood 1994), at least at radii beyond
$\sim~10^4R_{\rm S}$.

Second, the large orbital modulation of the NUV continuum intensity
(Fig.~4a) cannot be attributed to synchrotron emission from an ADAF
plasma, nor can it be explained by a symmetric accretion disk.
However, disk asymmetries are commonly observed, such as the bright
spot formed by the impact of the accretion stream on the outer edge of
the disk (Warner 1995).  We discuss and feature a second type of 
accretion disk asymmetry below.

The third argument for accretion-disk emission is the apparent
Planckian shape of the optical/UV continuum (Fig. 15), which resembles
the spectrum of a multi-temperature blackbody (\S4.2).  One motivation
for considering this specific model of the continuum are the results
obtained for J1118 during outburst, which showed in the low/hard state
the presence of a relatively cool accretion disk ($kT~\approx~24$~eV)
with a large inner disk radius ($R_{\rm tr}~\gtrsim~55R_{\rm S}$;
McClintock et al. 2001b; Esin et al. 2001).  The ADAF model, which
predicted the existence of this truncated disk in J1118, also predicts
that in quiescence the inner edge of the disk will have moved much
further out.

There are, however, two apparently serious objections to the accretion
disk model just described.  The first objection is that an effective
temperature of 13,000~K is far too high for a disk in quiescence; this
is a general problem for all models that seek to explain the
optical/UV emission of X--ray novae with an optically--thick accretion
disk model (Lasota, Narayan \& Yi 1996).  If the accretion outbursts
of X--ray novae are due to the dwarf nova instability as widely
assumed, then the effective temperature in quiescence cannot be much
greater than about 5000~K (Cannizzo 1993; Hameury et al. 1998).  Of course, the
radiation could come from hotter, optically-thin gas on the surface of
the disk --- the presence of strong emission lines in fact strongly
suggests this to be the case --- but then one does not expect the
standard multi-temperature blackbody model to apply, and it is no
longer straightforward to estimate $R_{\rm tr}$ from the observations.

The second objection or problem, one that has again been discussed in
the past (e.g., Lasota et al. 1996), is that a quiescent disk with a
small inner radius (e.g., $R_{\rm in}~\approx~1500R_{\rm S}$; \S4.2) 
cannot supply the mass accretion rate needed to
power the ADAF.  Lasota (2000) gives the following relation for the
maximum mass accretion rate $\dot M(R)$ in a quiescent disk at
radius $R$:
$$
\dot M(R) \approx 4.0\times10^{15} \left({M\over M_\odot}\right)
^{-0.88} \left({R\over 10^{10} ~{\rm cm}}\right)^{2.65}
~{\rm g\,s^{-1}}. \eqno (2)
$$
Setting $M=7M_\odot$ and $R=r_{\rm tr}R_{\rm S}$, where $r_{\rm tr}$ is the
transition radius between the disk and the ADAF, and assuming
that disk evaporation at radius $R_{\rm tr}$ feeds the ADAF, we have
$$
\dot M_{\rm ADAF} < 1.1\times10^{13} \left({r_{\rm tr}\over10^3}\right)
^{2.65} ~{\rm g\,s^{-1}}. \eqno (3)
$$
The ADAF models shown in Figure~13 all require mass accretion rates of
several times $10^{15} ~{\rm g\,s^{-1}}$
($\sim~10^{-10}~M_{\odot}$~yr$^{-1}$).  Clearly, this is incompatible
with a disk inner radius of $10^3R_{\rm S}$.  A radius of order 
$10^4R_{\rm S}$,
which is independently suggested by the H$\alpha$ line width, is much
more consistent with the ADAF model.  Some additional evidence for the
larger radius is provided by optical timing studies of quiescent X-ray
novae (Hynes et al. 2002; Zurita, Casares, \& Shahbaz 2003).

A motivation for elaborating a simple accretion disk model for J1118
that consists of a symmetric disk plus a bright spot (Warner 1995)
comes from a consideration of the NUV continuum light curve (Fig. 4a).
A bright spot is a bad model for explaining this light curve because
it predicts a maximum in intensity at just the orbital phase where the
deep minimum occurs.  On the other hand, this deep minimum at phase
0.7--0.8 (Fig. 4a) is very naturally explained as an absorption dip
due to matter in the accretion stream that is {\it not} stopped at an
impact region at the outer edge of the disk but overflows the disk
toward smaller radii.  Such UV and X--ray absorption dips at just this
phase have been observed in a number of systems and have been the
subject of intensive 3-D smoothed particle hydrodynamics (SPH)
simulations (Armitage \& Livio 1996; Hessman 1999; Kunze, Speith \&
Hessman 2001, and references therein).  These simulations show the
presence of substantial quantities of gas some $\sim~15-20^{\rm o}$
above and below the disk plane around orbital phase 0.7--0.8.  Thus
the presence of an absorption dip near this phase is expected in the
context of the stream--disk overflow model given the high inclination
of J1118, namely $i~\sim~80^{\rm o}$.

The SPH simulations and the observations of ``ordinary''quiescent
dwarf novae such as U~Gem and IP~Peg indicate that substantial stream
overflow is expected to occur for the mass accretion rates mentioned
above for the ADAF models (Kunze et al. 2001).  At these low mass
transfer rates, the vertical deflection of the gas at the outer edge
of the disk is relatively modest; the gas will move very quickly
inward with most of it striking the disk surface close to the
circularization radius (Kunze et al. 2001), $R_{\rm
circ}~\approx~3.9~\times~10^{4}R_{\rm S}$ (Frank, King, \& Raine
1992), where we have assumed q~=0.04 (Orosz 2001).  This radius is
$\sim4$~times the transition radius, $R_{\rm tr}$, adopted in \S4.1,
which is in harmony with the conclusions reached by Menou, Narayan, \&
Lasota (1999).  For convenience, we summarize here the various radii
discussed in this work: $R_{\rm in}~\approx~1500R_{\rm S}$, based on
the optical/UV continuum (\S4.2); $R_{\rm tr}~\lesssim~10^{4}R_{\rm
S}$, based on the width the H$\alpha$ line (\S4.1); $R_{\rm
tr}~=~10^{4}R_{\rm S}$, adopted in modeling the ADAF (\S4.1); and
$R_{\rm circ}~\approx~3.9~\times~10^{4}R_{\rm S}$.  For comparison,
the mean Roche lobe radius of the primary is $R_{\rm
L}~=~5.3~\times~10^{4}R_{\rm S}~=~1.6~R_{\odot}$.

There are important uncertainties in stream-disk overflow models, such
as the degree of cooling of the overflowing stream and the ratio of
the stream to disk scale heights at the disk edge (Armitage \& Livio
1998).  Nevertheless, a cold accretion disk ($T~\lesssim~5000~$K) plus
an overflowing accretion stream provides a rational, although
qualitative, explanation for the optical/UV results we have presented.
The broad, intense emission lines may be due to the gas in the
overflowing stream interacting with the disk over a range of radii and
Keplerian velocities, or they may arise in a chromosphere on the
surface of the cold disk. The hot, multi-temperature continuum source
($T~\sim~13,000~K$; \S4.2) may be attributed to emission generated by
the impact of the stream on the disk surface in the vicinity of the
circularization radius.  It is reasonable to expect that the impact of
the overflowing stream on the disk would produce a temperature of this
magnitude, since this is a typical blackbody temperature observed for
bright-spot emission from dwarf novae ($T~\approx~11,000-16,000~K$;
Warner 1995).  The UV continuum light from this hot, inner-disk region
is then modulated by the gas splashed up at the outer edge of the disk
near phase 0.7--0.8, thereby producing the NUV light curve (Fig.~4a).

This qualitative description of an accretion disk with an overflowing
accretion stream also provides a possible explanation for the small
and problematic disk radius, $R_{\rm in}~\approx~1500~R_{\rm S}$ that
we inferred from the optical/UV continuum spectrum (\S4.2).  Namely,
the normalization of the blackbody component does not give the
inner-disk radius, rather it is a measure of the heated area of the
disk in the vicinity of $R_{\rm circ}~\approx~3.9~\times~10^{4}_{\rm
S}$.  This hypothesis clears the way for concluding that the inner
disk radius is $R_{\rm in}~=~R_{\rm tr}~\sim~10^{4}R_{\rm S}$, as
indicated by the velocity width of the H$\alpha$ line (\S4.1) and as
required to power the ADAF (see the discussion below eq. 3).  Finally,
the N~V and Si~IV lines have high ionization potentials, 98~eV and
45~eV, respectively, compared to the ionization potentials of hydrogen
and Mg~II, 14~eV and 15~eV; possibly the N~V and S~IV lines are
generated in shocks near the circularization radius where the stream
impacts the disk

\section{Conclusions}

We have presented a multiwavelength spectrum of the black hole primary
in J1118 in its quiescent state at $L_{\rm
x}~\approx~4~\times~10^{-9}L_{\rm Edd}$ (Fig.~10). The results are
little affected by interstellar absorption or source variability.  The
following are the major conclusions that can be drawn from our
results:

\noindent
(i) The spectrum of J1118 does not appear to be peculiar to this
source or to this particular observation.  Rather, the very similar
spectrum of the quiescent black--hole A0620--00 indicates that we have
observed the canonical spectrum of a stellar--mass black hole radiating
at $L_{\rm x}~\sim~10^{-8.5}L_{\rm Edd}$.

\noindent
(ii) The X--ray component of the spectrum is well-fitted by a simple
power law, $\alpha~= 2.02~\pm~0.16$, with the column density fixed at
the value determined during outburst $N_{\rm
H}~\approx~1.2~\times~10^{20}$~cm$^{-2}$.  We are confident that the
X--ray emission comes from hot electrons near the black hole, which we
model as an ADAF that extends outward to the inner edge of an
accretion disk at $R_{\rm tr}~\sim~10^4~R_{\rm S}$.

\noindent
(iii) We ascribe the optical/UV component of emission to an accretion
disk that is truncated at its inner edge by the ADAF at $R_{\rm
tr}~\sim~10^4~R_{\rm S}$.  The disk model is strongly motivated by the
presence of broad Mg II and Balmer lines
(FWHM~$\approx~2400$~km~s$^{-1}$) in the spectrum; the shape of the
optical/UV continuum spectrum and the large-amplitude UV light curve
provide further motivation.

\noindent
(iv) The phase and large amplitude of the UV light curve can be
explained naturally by a UV-absorbing accretion stream that overflows
the outer edge of the disk. In this picture, the optical/UV continuum
($T~\approx~13,000~K$) is generated where the overflowing stream
strikes the disk surface in the vicinity of the circularization
radius, $R_{\rm circ}~\approx~3.9~\times~10^{4}R_{\rm S}$.

We conclude by comparing the feeble X--ray luminosity of J1118 to the
luminosities of other X--ray novae in quiescence, including systems
with neutron star primaries.  We do this by showing in Figure~16 an
update of the plot presented by Garcia et al. (2001) and recently by
Narayan et al. (2002).  For a detailed presentation of the data and
the motivations for plotting Eddington--scaled luminosity {\it vs.}
orbital period, see the above references.  Focusing on the
non--hatched region of this plot, the key conclusion is that black
hole X--ray novae are dimmer than neutron star X--ray novae by a
factor of 100 or more.  As discussed in Garcia et al. (2001) and
Narayan et al. (2002), the ADAF model provides a natural explanation
for this difference and hence strong evidence for the existence of
event horizons.  We should note that several authors have proposed
alternative explanations for Figure~16 that do not involve the presence
of an event horizon in black hole systems.  For a discussion and
critique of these proposals, see Narayan et al. (2002).

In response to the referee, in this paragraph we digress to defend
further our use of Eddington--scaled luminosities in comparing black
holes and neutron stars (Fig.~16).  Implicit in the above--stated
argument for the existence of event horizons is the assumption that
the mass transfer rates from the secondaries in the two kinds of
binaries are similar when expressed in Eddington units.  This
assumption is in fact reasonable.  For instance, King, Kolb \& Burderi
(1996) have estimated that the critical mass transfer rate $\dot
M_{\rm crit,irr}$ below which an X-ray binary with an irradiated
accretion disk would show transient behavior scales as
$$
\dot M_{\rm crit,irr} \approx 5\times10^{-11} \left({M_1\over
M_\odot}\right)^{2/3} \left({P_{\rm orb}\over 3~{\rm hr}}\right)
^{4/3} ~M_\odot{\rm yr^{-1}}.  \eqno (4)
$$
Since all the systems shown in Figure~16 are transients, they should all
lie below this critical mass transfer rate.  \footnote{King et
al. show that, for a magnetic braking model, most neutron star
binaries would lie above the line and would be persistent sources,
while most black hole binaries would be below the line and would be
transients.  This is irrelevant for our present argument since we
consider only those binaries in each category that are transients.}
Therefore, one should ideally compare the values of $L_{\rm
min}/(M_1/M_\odot)^{2/3}$ of neutron star and black hole binaries,
rather than $L_{\rm min}/(M_1/M_\odot)$, which is effectively what we
have done.  The difference in the two scalings is, however, rather
small compared to the very large luminosity difference that is seen
between neutron star and black hole systems in Figure~16.  Therefore, it
seems unlikely that the observed difference can be explained merely in
terms of different mass transfer rates in the two systems.  

We now return to our principal purpose in showing Figure~16, that is,
to compare the luminosity of J1118 to the luminosities of other X--ray
novae.  The figure is nearly identical to the one presented by Narayan
et al. (2002).  The main difference is that we have added the data for
J1118 and two other recently observed black hole systems: GRS~1009-45
(Hameury et al. 2002) and GRS~1124-683 (Sutaria et al. 2002).  The
data points for these three systems are labeled in Figure~16.
Restricting attention to black holes in the non-hatched region of the
plot and ignoring upper limits, one sees that J1118 and three other
systems (GRS~1009-45, GRO~J0422+32, and A0620--00) have
Eddington-scaled luminosities that are within a whisker of
$10^{-8.5}$.  Given the uncertainties in the distances to these
systems, not to mention the likelihood of source variability, the
quiescent luminosities of J1118 and these three systems are
indistinguishable.  The one surprising outlier is GRS~1124-683 which,
based on a single observation, is almost an order-of-magnitude more
luminous.

\acknowledgments

We thank the many people that made these simultaneous observations
possible and fruitful.  They include the {\it Chandra X--ray
Observatory} Director H. Tananbaum and the entire {\it Chandra} team
with special appreciation to M. Krauss, J. West and T.  Gokas for help
with the data analysis.  They include also the {\it Hubble Space
Telescope} team, in particular K. Peterson and H. Lanning, for
invaluable help in planning the STIS observations and James Davies and
others for help with the data analysis.  We also thank MMT Observatory
Director C. Foltz and his staff with special thanks to our telescope
operator A. Milone.  L. Hartmann and K. Luhman generously gave us the
three hours of high-quality MMT observing time that made possible the
radial velocity and spectral classification measurements of J1118.
R. Kurucz kindly provided the stellar model we used to derive the
non--stellar optical spectrum.  We gratefully acknowledge a discussion
on cataclysmic variable accretion disks with E. Schlegel, email
communications on the disk instability model with J. Cannizzo, and
comments on the manuscript from J. Miller.  We thank an anonymous
referee for a careful reading of our manuscript and helpful
comments. This work was supported in part by NASA contract NAS8-39073
to the {\it Chandra} X--Ray Center, contracts NAS8-38248 and
NAS8-01130 to the HRC Team, and grants GO-09282.01, NAG5-10813 and NAG
5-10780.

\newpage

\figcaption[fig1.eps]{The bars indicate the times of the X--ray ({\it CXO}),
UV ({\it HST}) and optical (MMT) observations.  The observations occurred on
two separate days, January 11 and 12 UT, which are designated {\it
epoch~1} and {\it epoch~2}, respectively.  The simultaneous
observations occurred during {\it epoch~2}; the dashed lines indicate
the strictly simultaneous coverage of the source.  The results
presented herein are comprised of all of the data obtained during {\it
epoch~2} plus the far--UV (FUV) data obtained with {\it HST} during
{\it epoch~1}.}

\figcaption[fig2.eps]{Composite UV spectrum of J1118 obtained
with STIS using the FUV/MAMA detector during {\it epoch~1} and the
NUV/MAMA detector during {\it epoch~2} (see Fig.~1).  The total
observation time in each band was 3.7~hr.  The dominant line in the
NUV band is the Mg~II doublet (2796 \AA; 2803~\AA).  The Si~IV doublet
(1394~\AA; 1403~\AA) is present in the EUV band.  The N~V~1240~\AA\
line is not shown here (see text).}
  
\figcaption[fig3.eps]{Evidence for the presence of the N~V and Si~IV
lines.  These coarsely binned spectra were derived for both lines as
follows: Using the summed 2-D spectral image (see text), we measured
the counts in a 10~x~10 pixel box that we centered on the spectrum and
advanced along the dispersion direction in 10-pixel steps, which
corresponds approximately to 5.8~\AA\ per step.  The background was
measured in parallel tracks that closely flanked the spectrum.  The
background level is indicated by the dashed line, which is the result
of a linear fit to the data plotted as open circles.  Based on just
the one high point in each spectrum, the statistical significance of
both the N~V and the Si~IV lines is $\approx4.5~\sigma$.  The
extremely intense geocoronal Ly$\alpha$ line, which is located to the
left of the vertical dotted line in (a), compromises the adjacent
background data and hence our assessment of the significance of the
N~V line.}

\figcaption[fig4.eps]{(a) Ultraviolet light curves {\it~vs.}
photometric orbital phase for (a) J1118 and (b) A0620--00 based on the
NUV data.  The intensities are normalized to the average intensity.
The solid dots correspond to the observed continuum intensity summed
over the bands 2250--2775 \AA\ and 2825--3000 \AA, which exclude the
Mg~II~2800~\AA\ line.  The small open circles correspond to the
intensity in the Mg~II line, which were determined by measuring the
intensity in the 2775--2825~\AA\ band and fitting the continuum in a
pair of adjacent bands, each of width 125~\AA.  The results shown here
do not take interstellar reddening into account.  Photometric phase
zero corresponds to conjunction with the secondary star in front of
the black hole.  The photometric phase of J1118 was obtained by adding
a quarter of an orbital cycle to the spectroscopic phase computed
using the time of maximum velocity (sect. 2.4.1), and the phase of
A0620--00 was taken from the ephemeris of Gelino, Harrison, \& Orosz
(2001).  We note that the interpretation of the continuum light curve
of A0620--00 (b) is problematic because its minimum does not occur
near phase 0.7, but it is especially so because the inclination of
A0620--00 is low: $i~=~41~\pm~3^{\rm o}$ (Gelino et al. 2001).  The
stream-disk overflow model requires $i~>~65^{\rm o}$ (Kunze et
al. 2001).}

\figcaption[fig5.eps]{Ultraviolet light curves {\it~vs.} time for
J1118.  The intensity data are identical to that plotted in Figure~4a.
Note that this version of the light curve is irregular compared to the
regular variations seen in Figure~4a.}

\figcaption[fig6.eps]{Spectroscopic data folded on the orbital period
given in the text. (a) Radial velocity measurements
of the secondary star.  The smooth curve is a fit to a circular orbit
based on the phase (see text), velocity amplitude
($K~=~684~\pm~15$~km~s$^{-1}$), and systemic velocity determined using
these data. (b) The residual differences between the data and the
fitted curve.  As shown, spectroscopic phase zero (photometric phase
0.25) corresponds to the time of maximum redshifted velocity of the
secondary.}

\figcaption[fig7.eps]{(a) The spectrum of J1118 in the
rest-frame of the secondary star exhibiting numerous spectral
features characteristic of a mid-K dwarf. (c) The spectrum of the
best-matching template star, GJ563.1. (b) The spectrum of J1118 minus
the spectrum of GJ563.1.  The subtraction removes the K-star lines
quite effectively except in the vicinity of the Mg~b complex near
5175~\AA.  The prominent, broad emission feature in the difference is
due to He~I~5875~\AA\ disk emission, which is observed in other
quiescent X--ray novae (e.g., Orosz et al. 1996).}

\figcaption[fig8.eps]{Low-resolution optical spectrum of J1118
corrected for reddening. (a) The observed spectrum.  H$\alpha$ and
three higher-order Balmer lines are the most prominent features.  Also
evident is the blended complex of absorption lines near
Mg~b~5180~\AA. The spectral resolution, which was seeing-limited, is
about 12~\AA\ (FWHM). We estimate that the uncertainty in the fluxes
is 0.1~mag. (b) The spectrum of the non--stellar or residual emission
obtained by subtracting a Kurucz K5V model spectrum from the spectrum
above.  This non--stellar spectrum has been boxcar-smoothed to 20~\AA\
resolution (FWHM).  We estimate that the uncertainty in the fluxes are
0.1~mag at the blue end, increasing monotonically to 0.3~mag at the
red end.}

\figcaption[fig9.eps]{RMS difference between the rest--frame spectrum
 of J1118 and various template spectra over a range of spectral types
 and metallicities.  Each of the normalized and resampled template
 spectra were scaled by various values of a weight factor $w$ between
 0.0 and 1.0 in steps of 0.02 and subtracted from the rest-frame
 spectra of J1118.  The scatter in each difference spectrum was
 measured by fitting a low-order polynomial and computing the rms
 difference (for further details, see Orosz et al. 2002).  The minimum
 value of this difference is plotted here for 20 observations of the
 16 template stars.  The results indicate that the secondary is near
 K5 and probably metal-poor.}

\figcaption[fig10.eps]{Composite multiwavelength spectrum of J1118
corrected for reddening.  The MMT optical spectrum on the far left of
the plot is identical to the spectrum shown in Figure~8b (see caption
for an estimate of uncertainties).  The STIS NUV spectrum, punctuated
by its intense Mg~II line, appears at somewhat higher frequency,
log($\nu)~\sim~15.05$ (cf. Fig.~2).  Next is the FUV spectrum, which
is centered at log($\nu)~\sim15.3$ and plunges downward to
log($\nu$$F_{\rm \nu})~-14.1$.  We estimate that the uncertainties in
the NUV and FUV fluxes are $\approx~0.1$~mag.  Finally, the best-fit
X--ray model for the data is indicated by the heavy, horizontal line.
The 90\% confidence error box is defined by the flanking curved
lines.}

\figcaption[fig11.eps]{Same as Figure~10 except with {\it HST} and {\it
Chandra} data for A0620--00 superimposed.  The NUV spectrum for A0620--00
is a rebinned version of published data (McClintock \& Remillard
2000).  For the derivation of the X--ray spectrum and error box for
A0620--00, see \S2.1.2.}

\figcaption[fig12.eps]{Results for a grid of ADAF models of XTE
J1118+480 with different choices of $p$ and $\delta$.  Filled dots
represent models which agree with the X--ray data.  The corresponding
spectra are shown in Figure~13.  Crosses represent models which do not
agree with the data.  The spectra of two of these models are shown in
Figure~14.}

\figcaption[fig13.eps]{X--ray spectra of the five models in Figure~12
that agree with the X--ray data on XTE J1118+480.  The models
correspond to $p, \delta$ = 0.2, 0.001 (solid line), 0.4, 0.1 (dotted
line), 0.6, 0.3 (short dashed line), 0.6 0.4 (long dashed line), and
0.8, 0.7 (dot-dashed line).}

\figcaption[fig14.eps]{X--ray spectra of two representative models that
do not agree with the X--ray data.  The models correspond to $p,
\delta$ = 0, 0.1 (solid line) and 0.8, 0.1 (dashed line).}

\figcaption[fig15.eps]{A blowup of the optical/UV spectrum of J1118,
which is shown in Figure~10.  From left to right, the dominant emission
lines in each band are $H\alpha$ in the optical, Mg~II in the NUV and
Si~IV in the FUV (cf. Figures~2~\&~8).  (a) The curves superimposed on
the data are for simple disk blackbody models (see text), which have
been normalized to match the flux at log($\nu)~=15.05$, where the
observations are most secure.  The solid curve with $kT$~=~1.1~eV
appears to best conform to the observations.  (b) The curves
correspond to single-temperature blackbody models with different
temperatures.  None of the models matches the data.}

\figcaption[fig16.eps]{Quiescent luminosities of black hole X--ray
novae (filled circles) and neutron--star X--ray novae (open circles)
plotted {\it vs.} the orbital period.  Only the lowest quiescent
detections or {\it Chandra/XMM} upper limits are shown.  The
non--hatched region includes both black--hole and neutron--star
systems and allows a direct comparison between the two classes of
X--ray novae.  The figure is identical to the one presented by Narayan
et al. (2002) except that (1) for Aql~X--1 we have plotted the lower
luminosity given in their Table~2 and (2) the new data for J1118,
GRS~1009--45 and GRS~1124--683 (Nova Mus 1991) have been added.}  

\newpage
\begin{figure}
\figurenum{1} 
\plotone{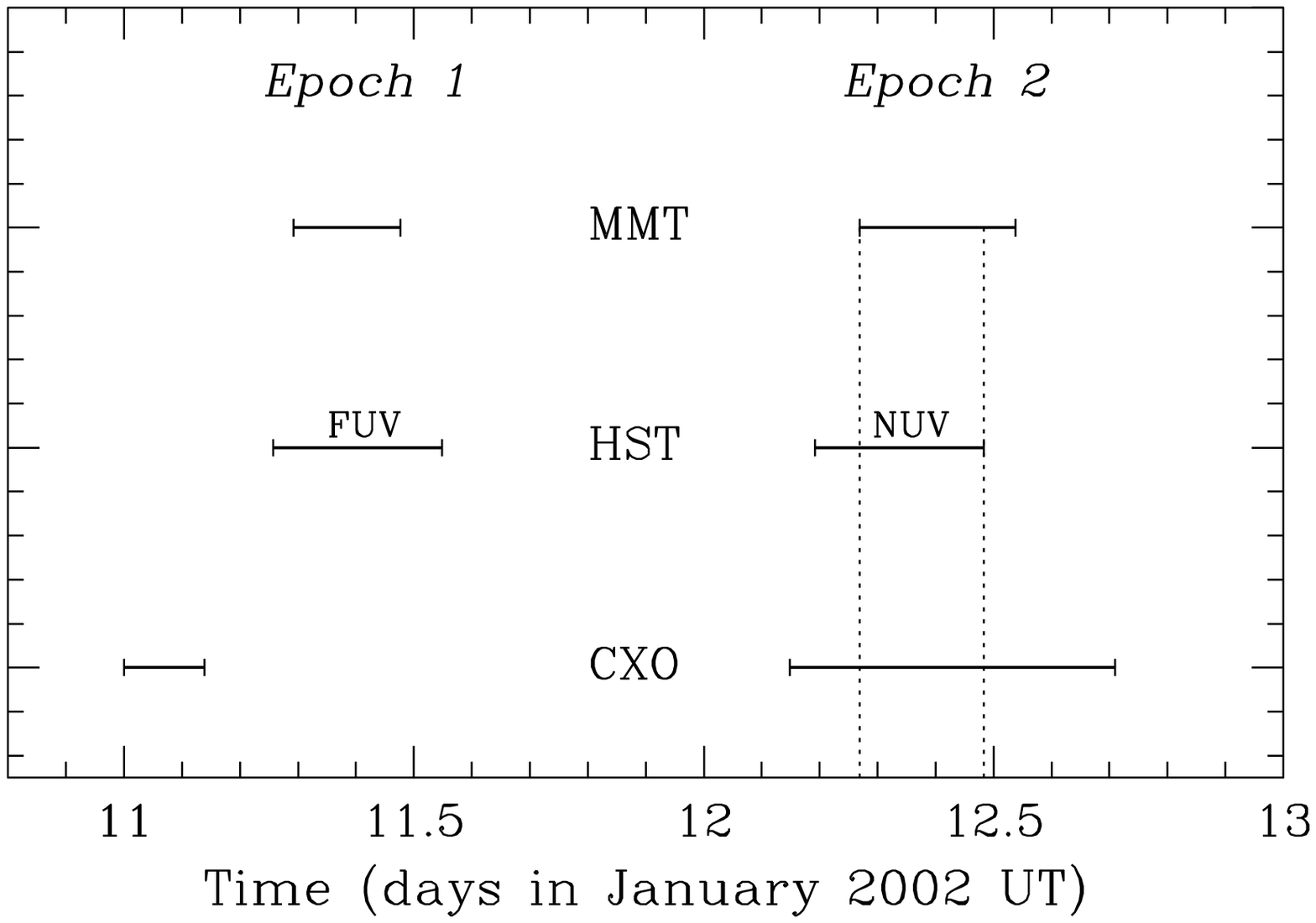}
\caption{ }
\end{figure}

\newpage
\begin{figure}
\figurenum{2}
\plotone{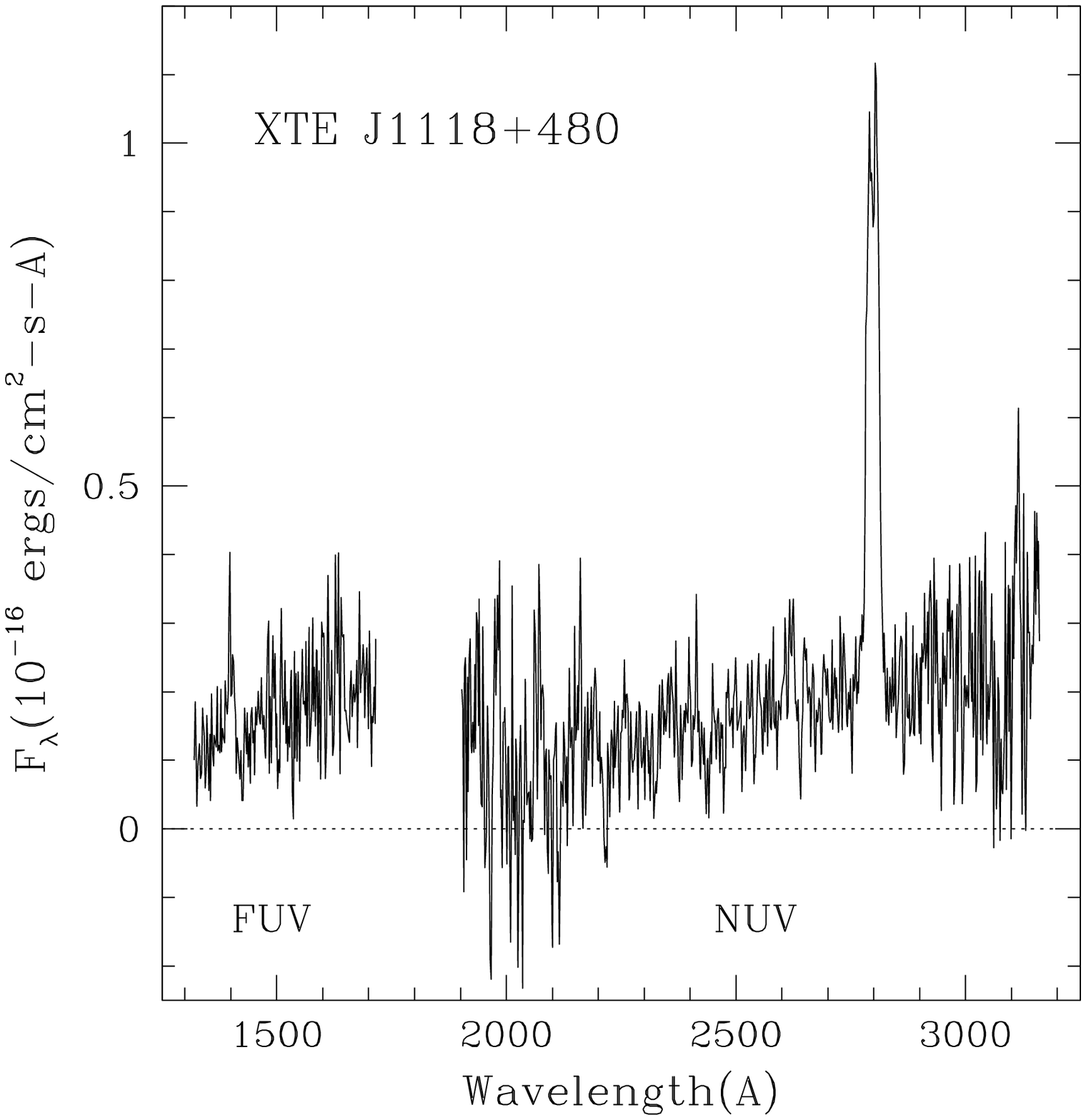}
\caption{ }
\end{figure}

\newpage
\begin{figure}
\figurenum{3}
\plotone{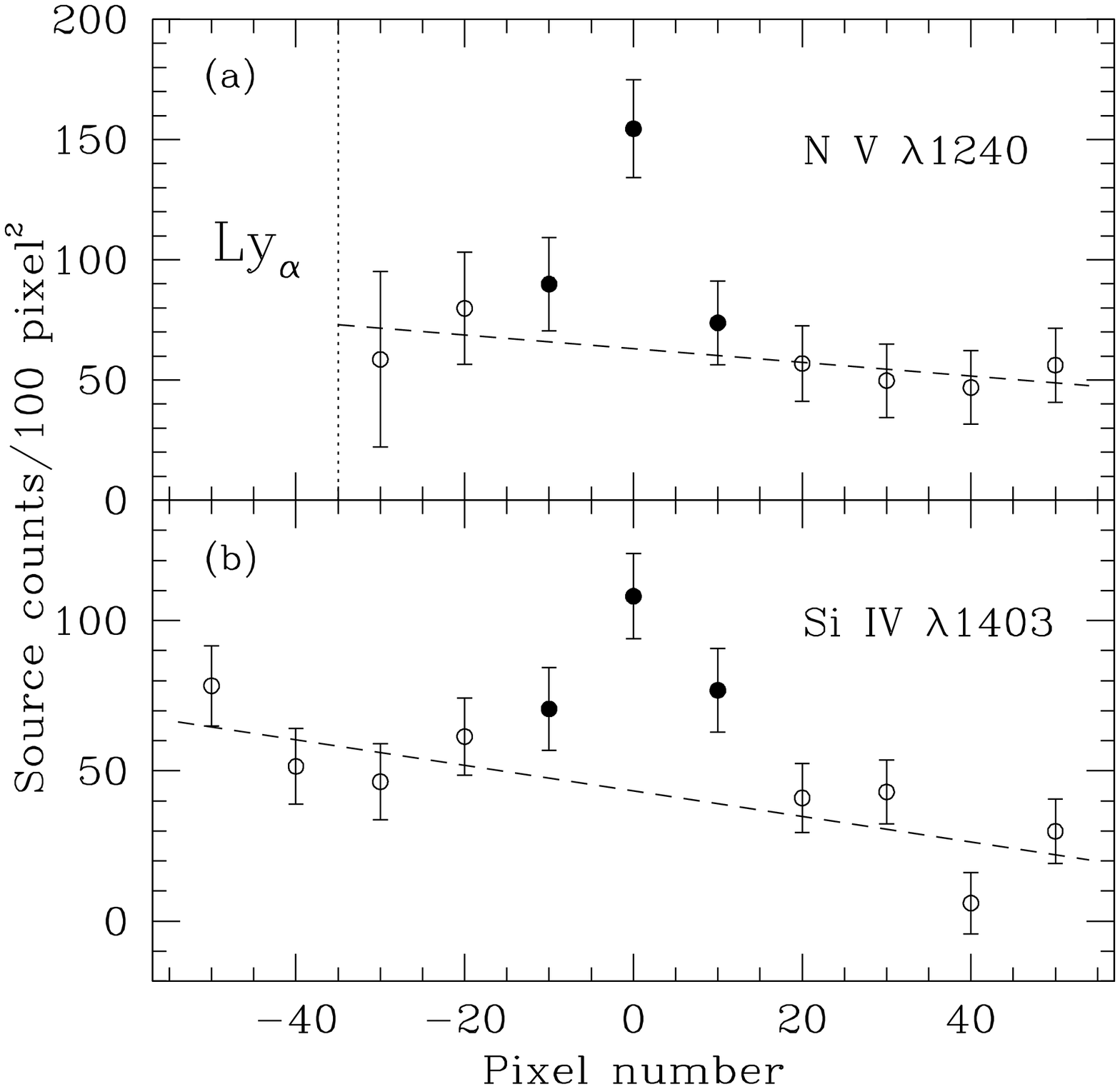}
\caption{ }
\end{figure}

\newpage
\begin{figure}
\figurenum{4}
\plotone{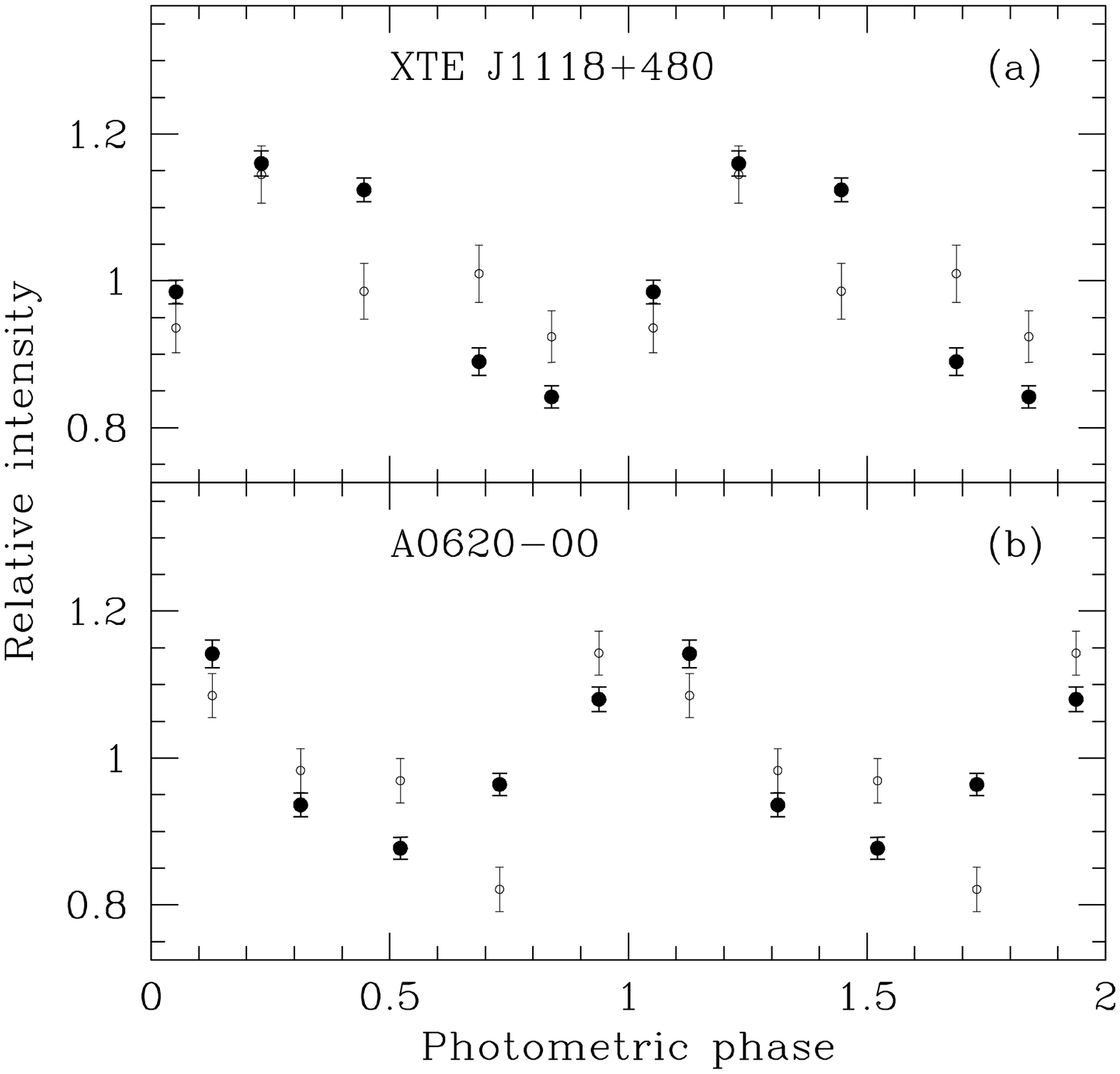}
\caption{ }
\end{figure}

\newpage
\begin{figure}
\figurenum{5}
\plotone{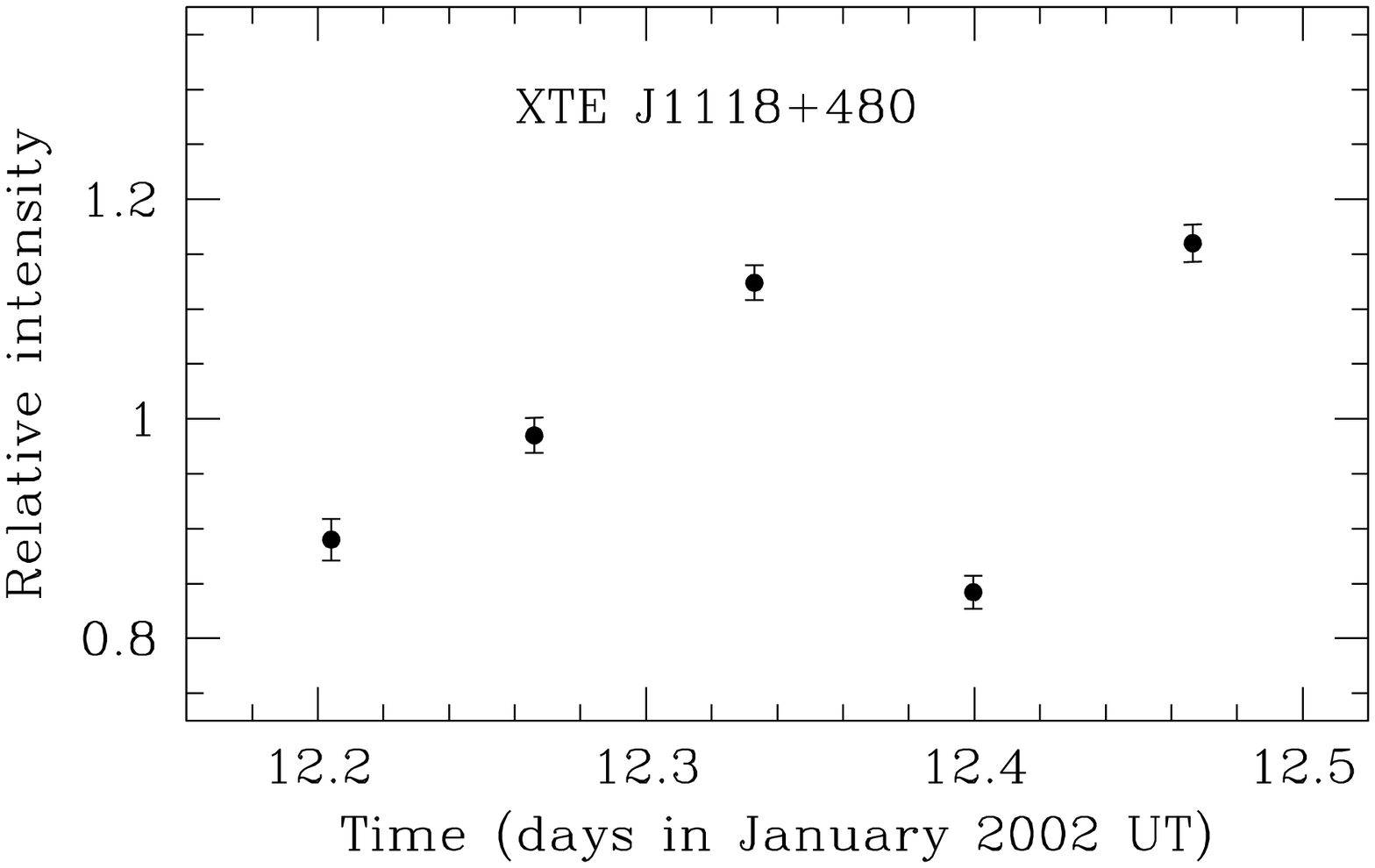}
\caption{ }
\end{figure}

\newpage
\begin{figure}
\figurenum{6}
\plotone{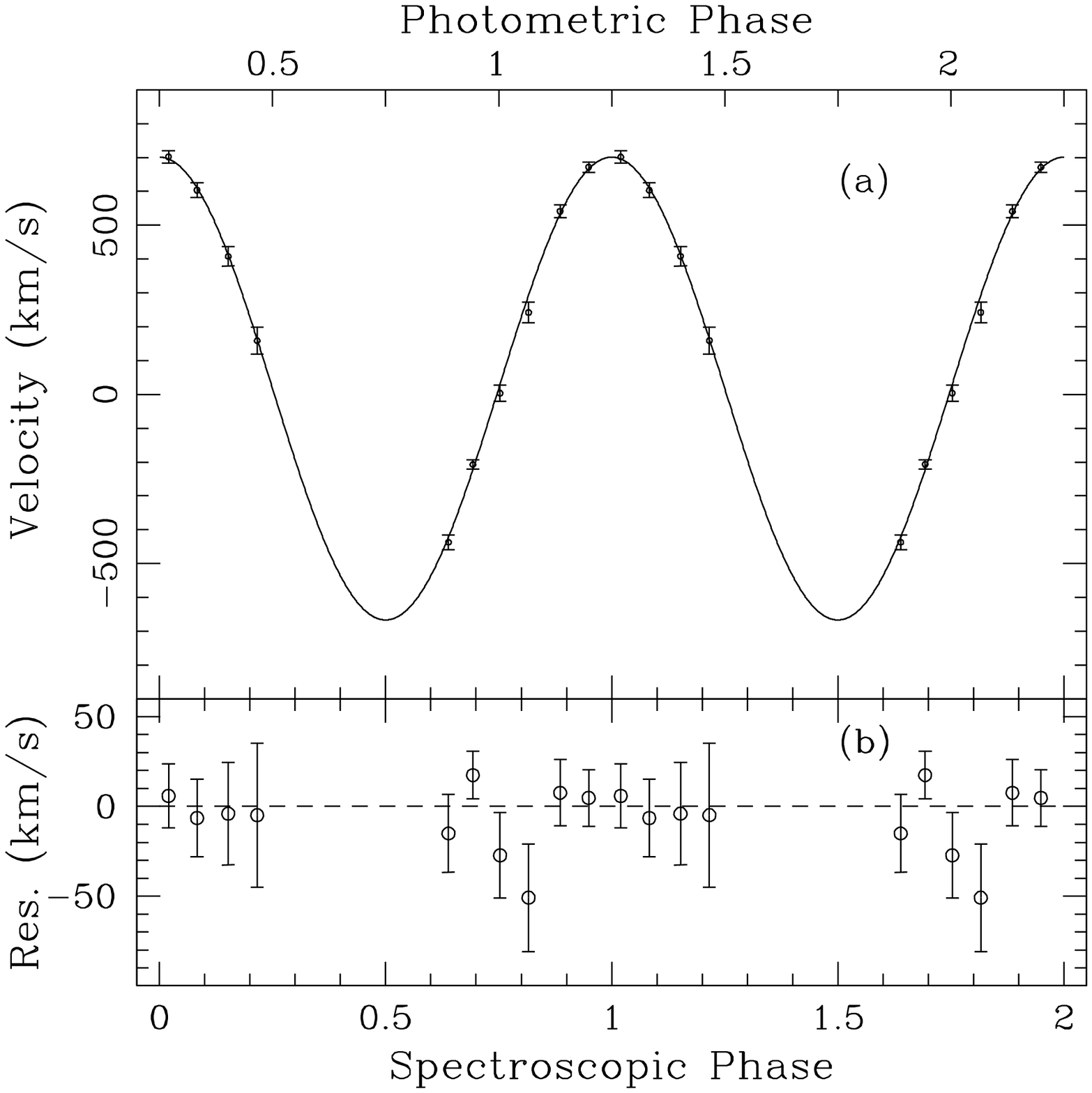}
\caption{ }
\end{figure}

\newpage
\begin{figure}
\figurenum{7}
\plotone{f7.eps}
\caption{ }
\end{figure}

\newpage
\begin{figure}
\figurenum{8}
\plotone{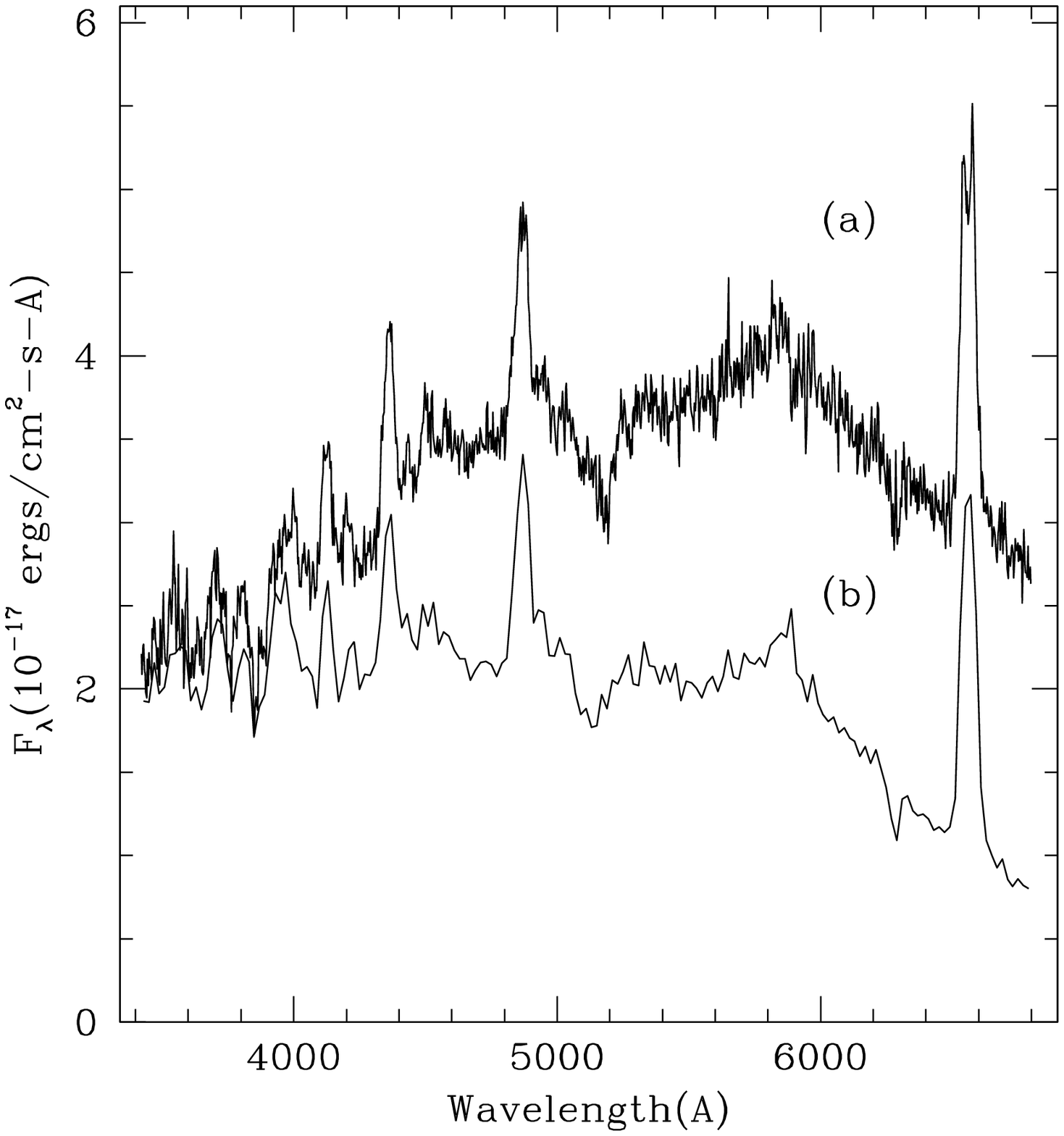}
\caption{ }
\end{figure}

\newpage
\begin{figure}
\figurenum{9}
\plotone{f9.eps}
\caption{ }
\end{figure}

\newpage
\begin{figure}
\figurenum{10}
\plotone{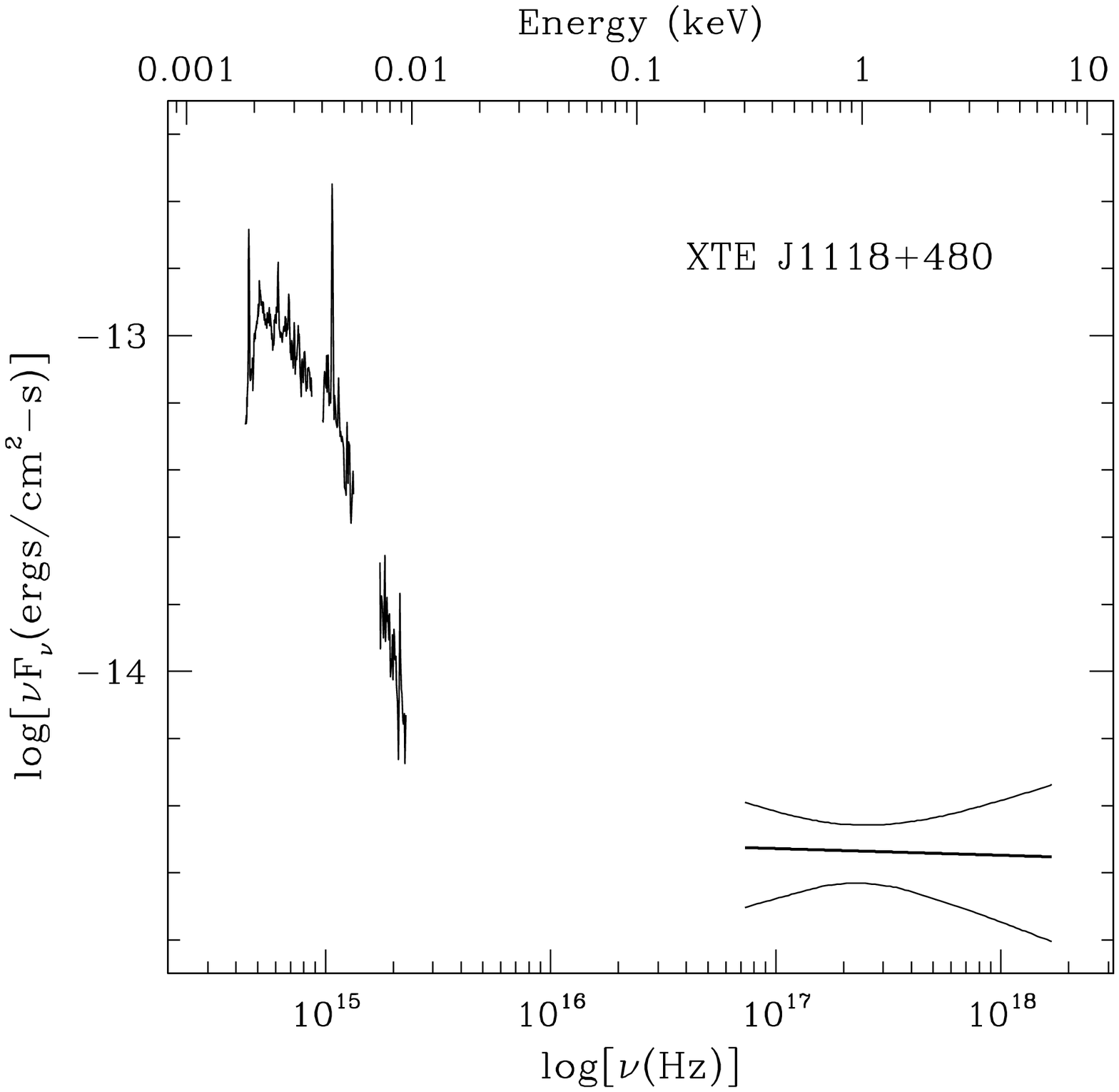}
\caption{ }
\end{figure}

\newpage
\begin{figure}
\figurenum{11}
\plotone{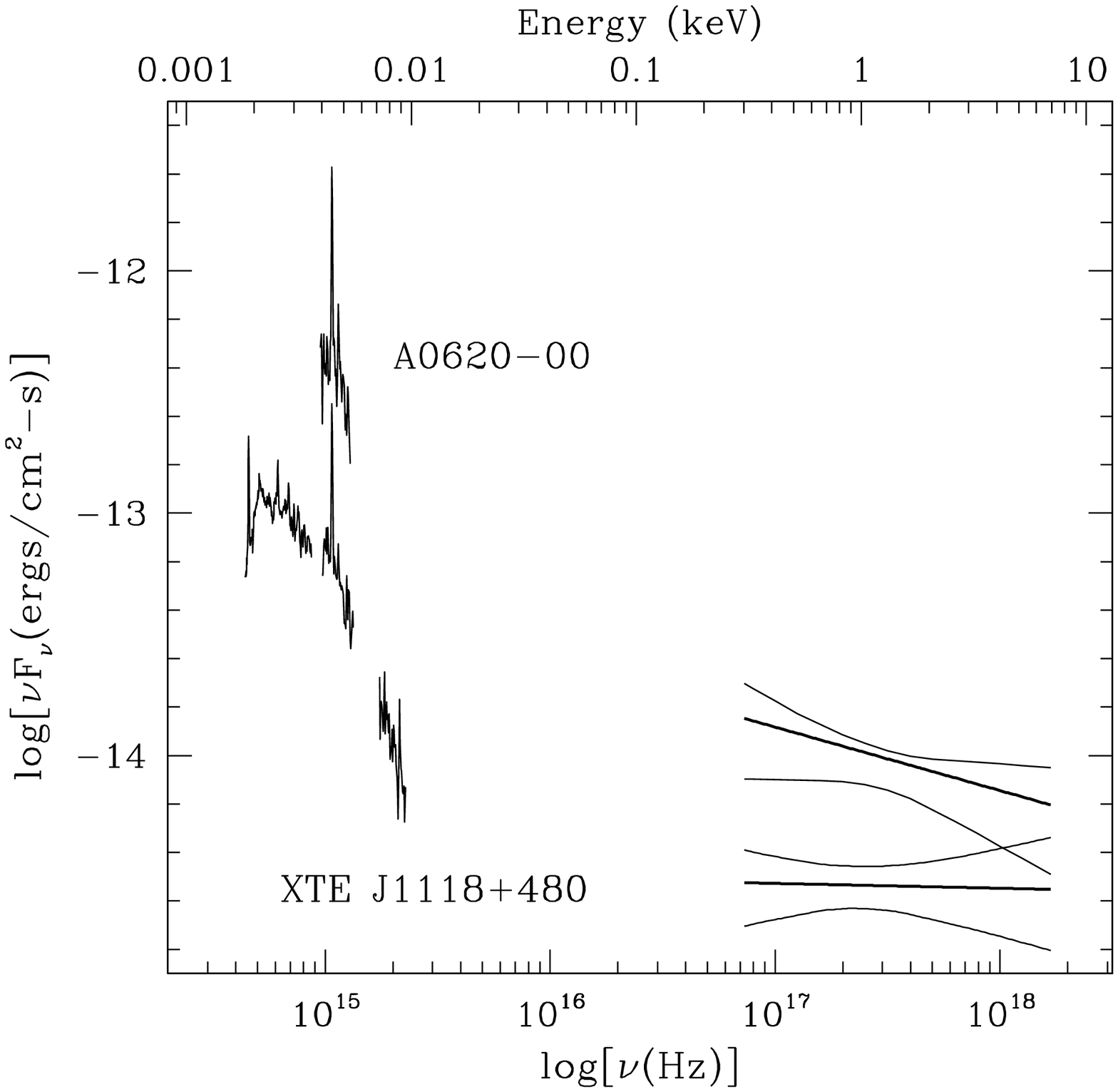}
\caption{ }
\end{figure}

\newpage
\begin{figure}
\figurenum{12}
\plotone{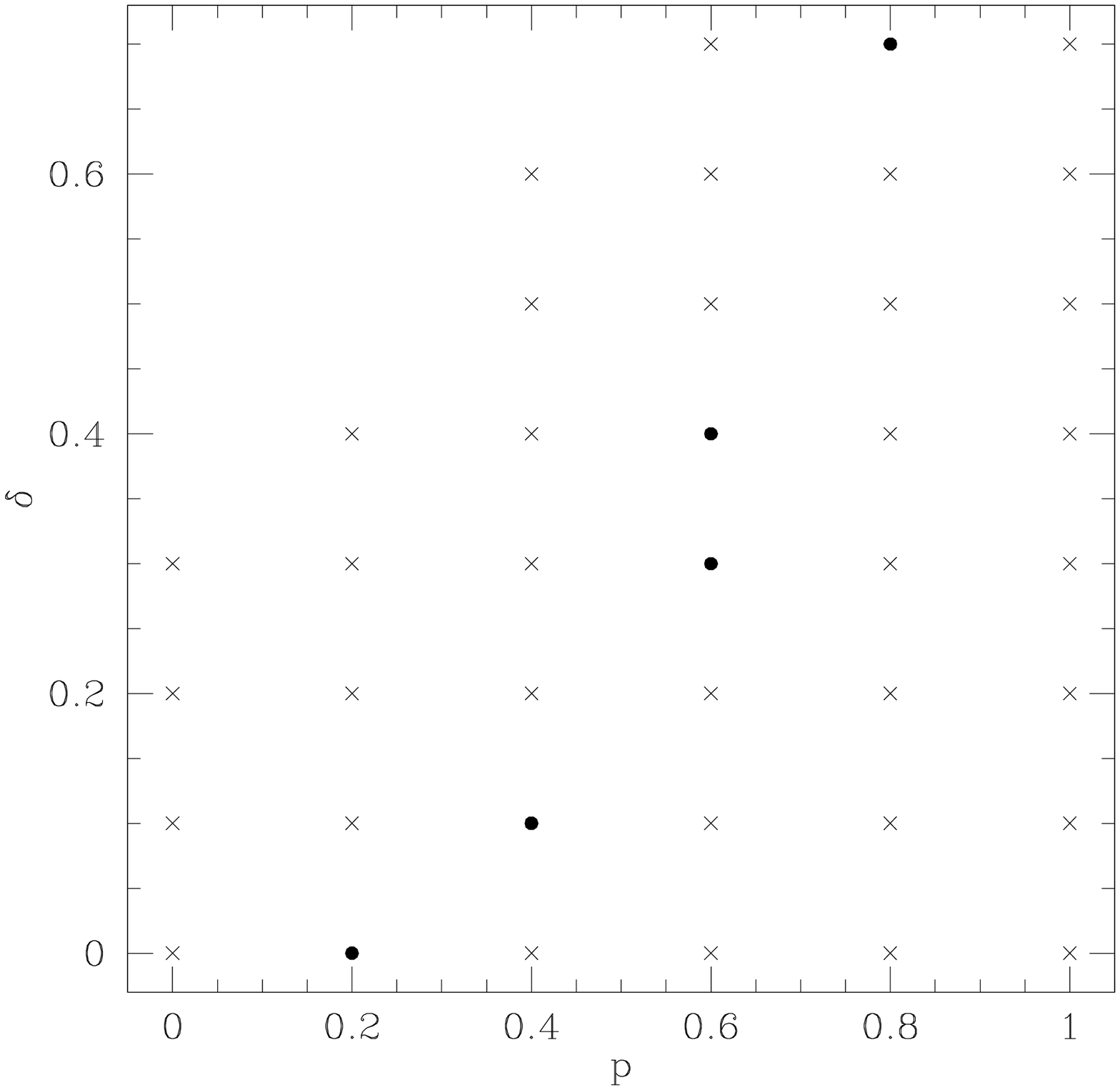}
\caption{ }
\end{figure}

\newpage
\begin{figure}
\figurenum{13}
\plotone{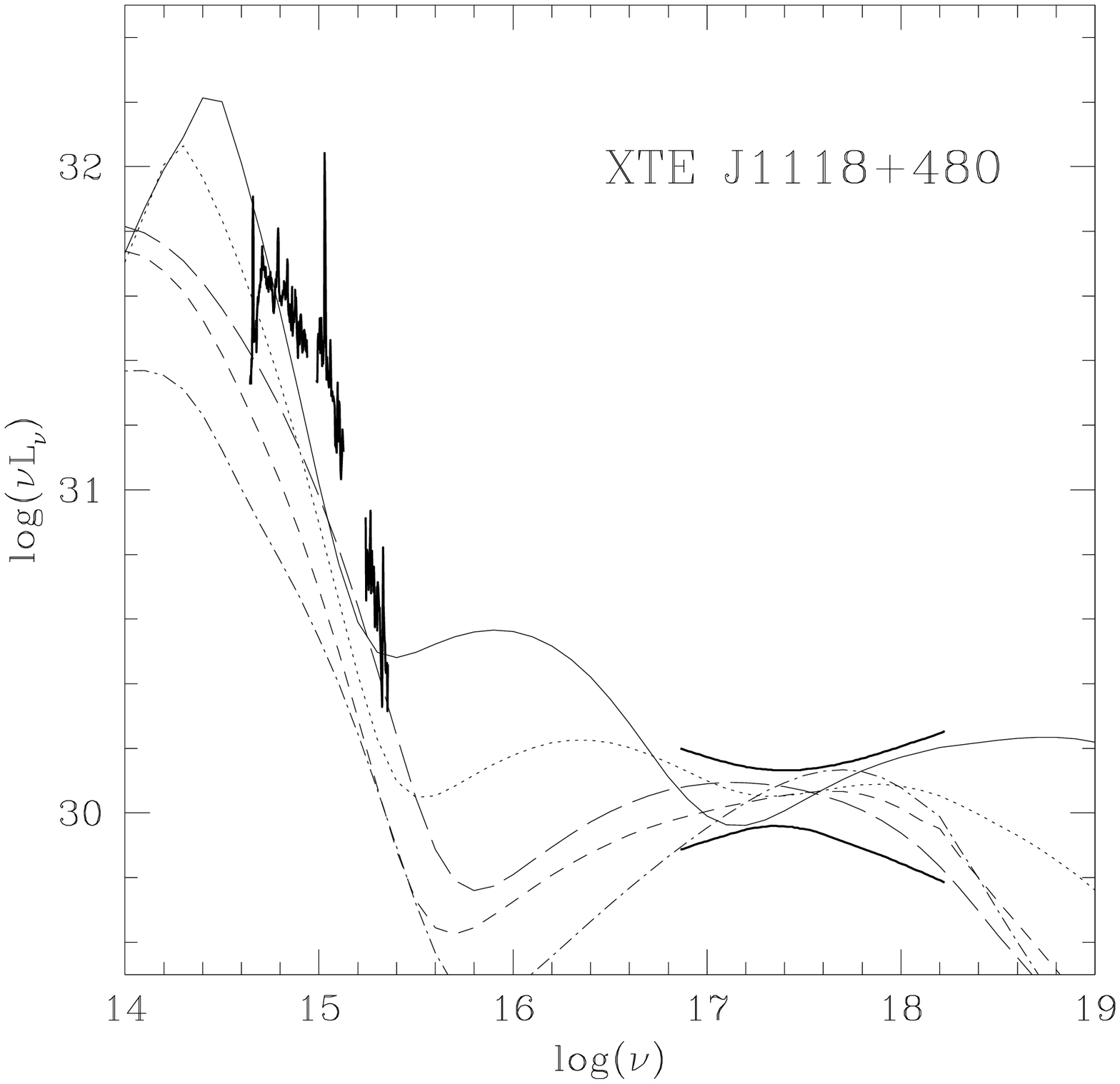}
\caption{ }
\end{figure}

\newpage
\begin{figure}
\figurenum{14}
\plotone{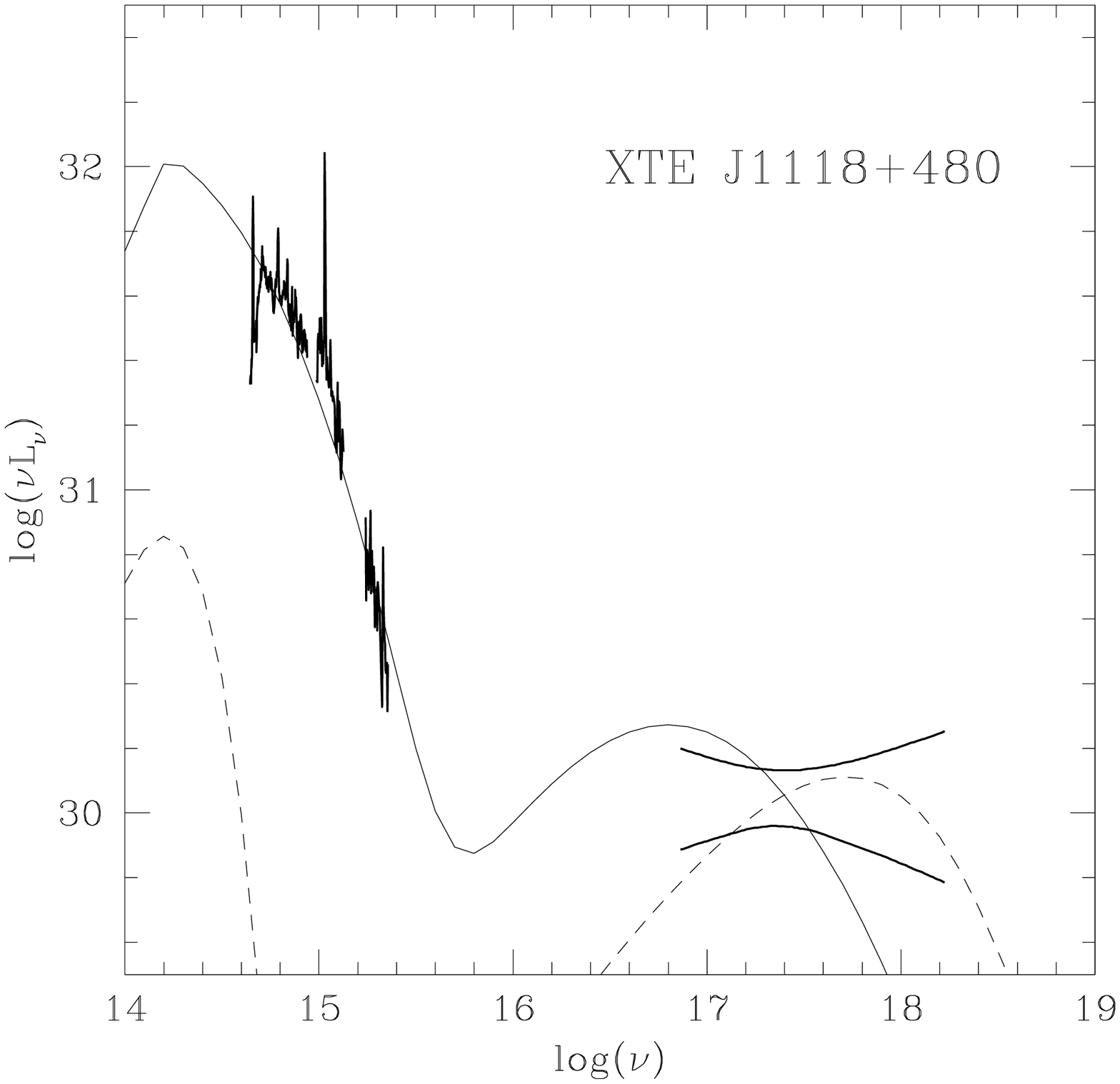}
\caption{ }
\end{figure}

\newpage
\begin{figure}
\figurenum{15}
\plotone{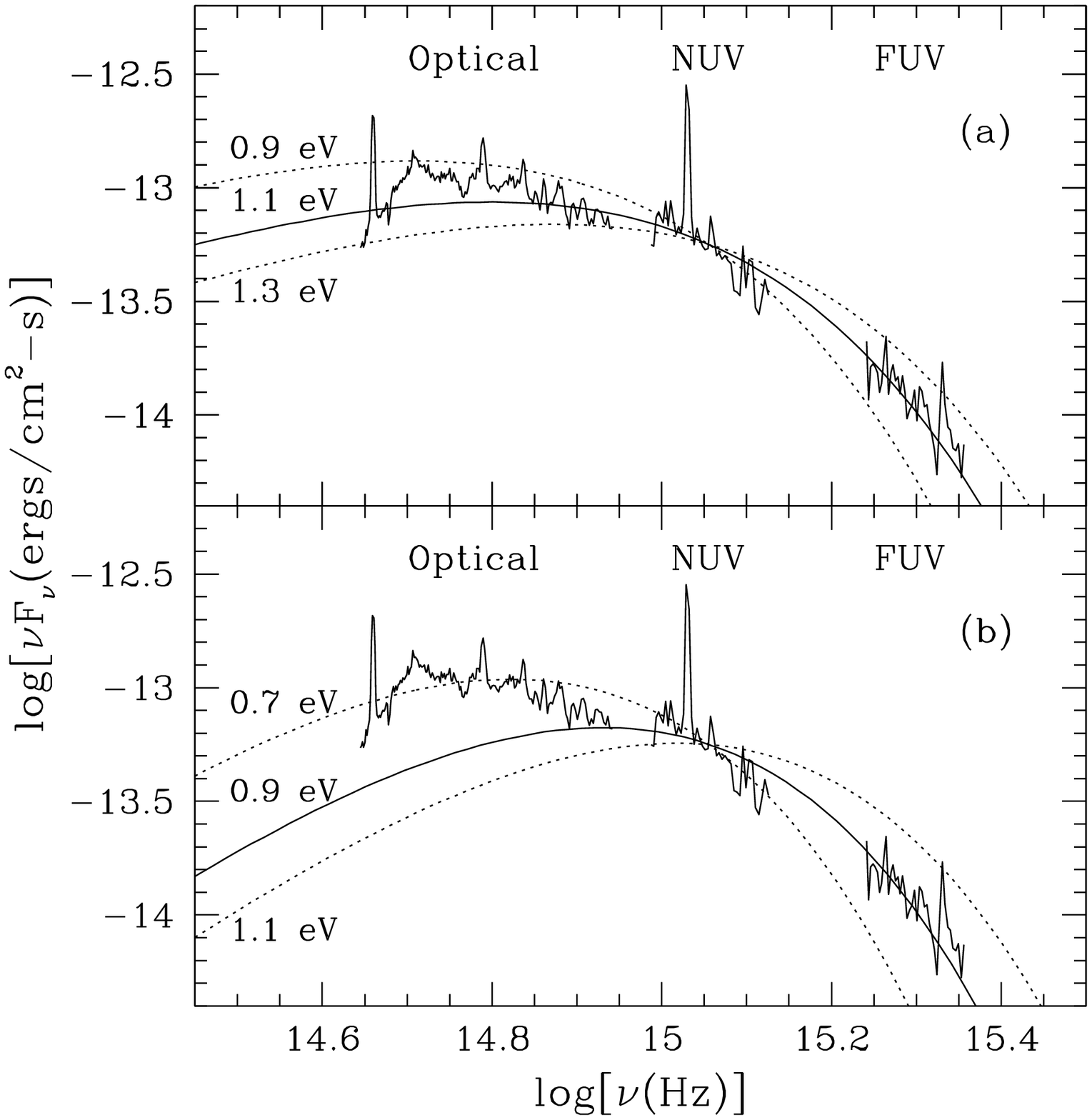}
\caption{ }
\end{figure}

\newpage
\begin{figure}
\figurenum{16}
\plotone{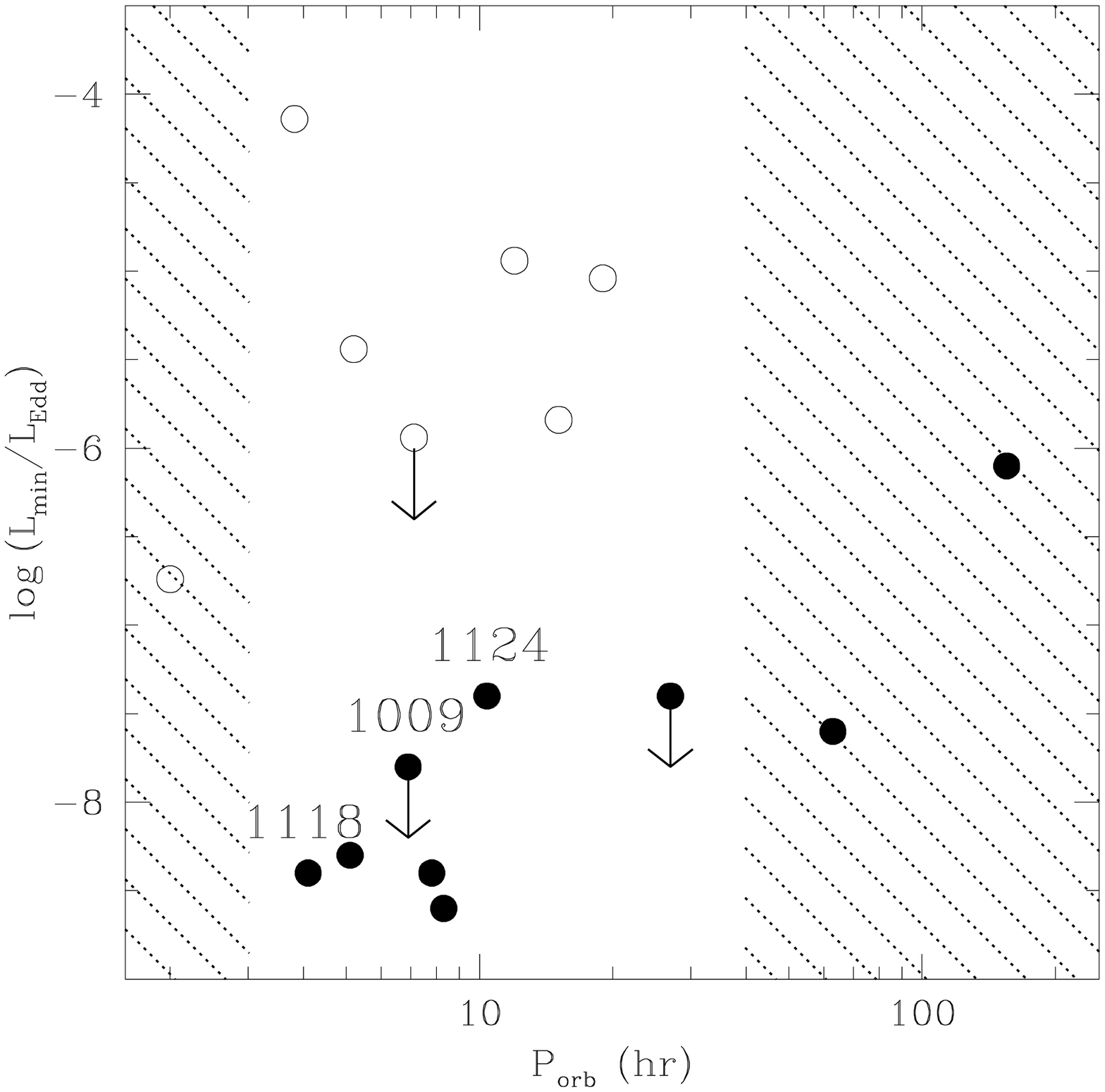}
\caption{ }
\end{figure}

\end{document}